# Model predicts fundamental role of biomechanical control of cell cycle progression during liver regeneration after partial hepatectomy

Stefan Hoehme, Rolf Gebhardt, Jan G. Hengstler, Dirk Drasdo

## Abstract

Partial hepatectomy (PHx) is a surgical intervention where a part of the liver is removed. Due to its extraordinary capacity to regenerate, the liver is able to regenerate about two-thirds of its mass within a few weeks. Nevertheless, in some patient's regeneration fails. Understanding the principles and limitations underlying regeneration may permit to control this process and prospectively improve the regeneration. Here, we established a simulation model to mimic the process of regeneration in the liver lobe of a mouse. This model represents each hepatocyte individually and builds upon a previous computational model of regeneration of drug induced damage in a single liver lobule. The present study simulates entire liver lobes that consist of hundreds to thousands of lobules. It accounts for biomechanical control of cell cycle progression (Biomechanical Growth Control), which has not been considered in that previous work. The model reproduced the available experimental observations only if BGC was taken into account. Interestingly, the model predicted that BGC minimizes the number of proliferating neighbor cells of a proliferating cell resulting in a checkerboard-like proliferation pattern. Moreover, the model predicted different cell proliferation patterns in pigs and mice that corresponded to data obtained from regenerating tissue of the two species. In conclusion, the here established model suggest that biomechanical control mechanisms may play a significant role in liver regeneration after PHx.

## Introduction

Due to its exposure to toxic compounds, the liver evolutionary acquired an extraordinary capacity for self-renewal [29]. The high regeneration capacity allows the partial surgical removal of the organ, partial hepatectomy (PHx), as a therapy of neoplasms, intrahepatic gallstones or cysts, whereby the remnant liver regenerates its mass within three to 10 days, depending on the species. However, the precise mechanisms and underlying principles controlling liver regeneration are still not fully understood and in some patients the regeneration process following PHx fails.

The liver consists of repetitive functional and anatomical building blocks, so called liver lobules. A human liver consists of about a million lobules, while a mouse liver comprises only several thousand lobules. Blood from the intestine reaches the liver via the portal vein. Moreover, it is supplied by arterial blood from the liver artery. Both, blood from the portal vein and from arterial vein branches reach the periphery of the liver lobules, from where it is drained through a network of micro-vessels, named sinusoids, towards the central vein located in the center of the lobule. Liver lobules of the mouse have mean diameters of about 500 micrometers perpendicular to the orientation of the central vein. In mice, the liver lobules are organized in five lobes, each encapsulated by a layer of connective tissue, the Glisson capsule. Liver regeneration after PHx is characterized by a massive increase of the cell mass followed by a remodeling phase. During the mass recovery phase that was addressed in the present study, proliferating hepatocytes have to push their neighboring cells to generate space for the daughter cells. In principle, this may lead to an increase of pressure and unphysiological compression. If 66% of the liver mass is resected (2/3 PHx), the remaining lobes have to increase their volume by a factor of three. Under the simplified assumption that liver lobes are spherical, an increase of 66% in volume corresponds to an increase of the lobe diameter to $3^{1/3}$ of its original value, corresponding to 44%. This increase occurs against the resistance of the Glisson capsule. This means that proliferating cells in the interior of a lobe need to generate a net force that is high enough to cause a net displacement of the lobe border by about 44% of the lobe diameter.

The situation is conceptually reminiscent of recent experiments of growing multicellular spheroids in elastic alginate capsules [1]. Here, multicellular spheroids growing within elastic alginate capsules reduce their expansion speed significantly when they touch the capsule demonstrating an influence of mechanical stress on cell cycle progression. Comparing the remaining speed of expansion and the mechanical resistance of the alginate capsule, it was possible to infer the influence of mechanical stress on individual cell cycle progression (Van Liedekerke et. al., 2019). Mechanical stress on cells has been observed to affect cell cycle progression also in other experimental situations [13, 18, 27, 30, 42], and it may impact on tissue form [25, 34]. How mechanical stress affects growth and form has also been studied in numerous model approaches [2, 6, 10, 11, 52].

While it can now be considered undeniable that sufficiently high mechanical compressive stress on a cell affects its growth and division, it remains unclear whether the cell senses the compression of its volume or directly the pressure exerted on it.

The above observations raise the question whether mechanical stress may build up in the liver lobe as a consequence of the growth during regeneration after PHx, and whether this may critically influence the regeneration process. To address this question, we here developed a computational model for liver regeneration after PHx by studying the growth of a single lobe down to the level of micro-architecture. The model was parameterized from biological experiments after ⅔ PHx and a cell-based model of liver regeneration after intoxication by carbon tetrachloride [32]. Cell-based (also named "agent-based" or "individual-cell"- or "single-cell"-based) models attract increasing interest to mimic multicellular processes [3, 36, 52] as they represent a direct approach of systems as single-cell resolution and permit to straightforwardly include intracellular mechanisms [33, 35, 46, 46]. Such models display each individual hepatocyte in a virtual liver lobule constructed from confocal laser scanning micrographs and hence represent the micro-architecture of liver tissue [22].

Recently, significant effort has been made towards mathematical models on blood or lymph flow, molecular transport, or metabolism in liver [4, 5, 7, 8, 15–17, 28, 37, 38, 45, 47–49]. These models consider the lobule level within schematic, regular hexagonal geometries or within compartment models without spatial representation of microarchitecture. The present model focuses on cells as individual basic modeling units organized in space in a realistic tissue microarchitecture and can hence be regarded as an in-silico abstracted copy of the real system.

In this study, we developed a model of liver regeneration after PHx that comprises an entire lobe consisting of numerous lobules (Fig.1). In a first step the model was used to explain liver regeneration in mouse. Besides cell kinetic parameters the model addresses biomechanical aspects. In particular the possible role of a biomechanical growth control on cell cycle progression, thereafter abbreviated as BGC was investigated. In a next step the model was extrapolated from mouse to pig by re-adjustment of the model parameters. The underlying question was whether the model could help in an extrapolation from mouse to humans. Livers from pigs were studied, because their size is in the same order of magnitude as that of humans, while livers of mice are about three orders of magnitude smaller. Moreover, it is difficult to obtain human liver tissue at defined periods after PHx, while such material is available from pigs.

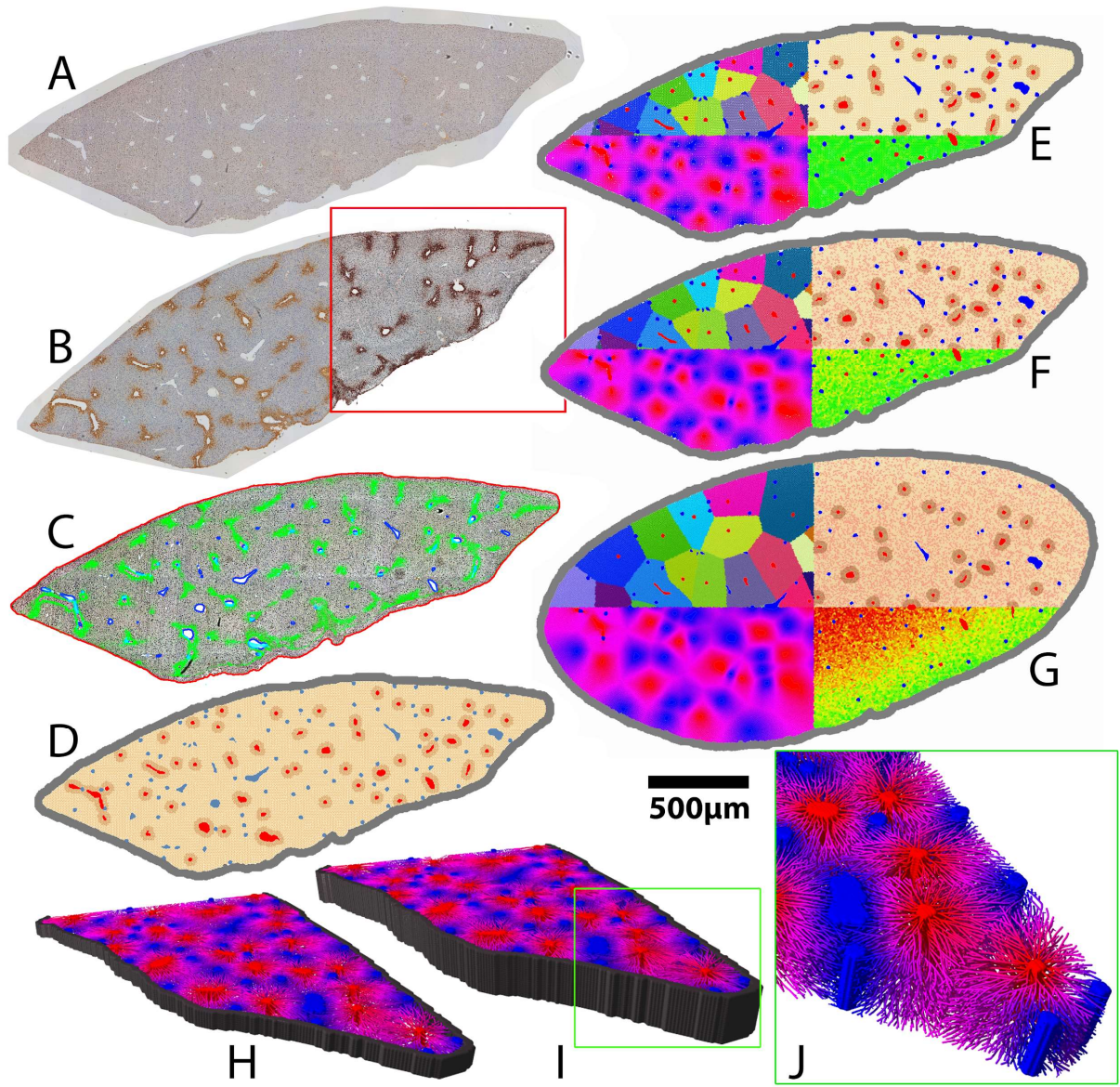

**Figure 1**: Construction of a single-cell based model from two whole slide scans of a liver lobe (A+B) by an image processing and analysis chain. (A) PCNA stained micrograph of a mouse liver lobe. (B) A neighboring slice stained for glutamine synthetase (GS). (C) Intermediate step in which the contrast of the micrographs was enhanced by contrast limited adaptive histogram equalization (CLAHE). The localization of the capsule (red outline) and of larger vessels (blue/cyan) was determined. The effect of CLAHE is illustrated within the red rectangle in (B). GS permits to distinguish between central veins and portal veins or arteries. Each central vein of a liver lobule is circumvented by GS positive hepatocytes. The green coloring shows the GS staining of (B) used to identify the central veins among the larger vessels in the image (red in (D)). E-G: Growth simulation of a liver lobe. Time series of proliferating and growing lobe. (E) t = 0 days, (F) t = 2 days, (G) t = 5 days. H-J: Exemplary 3D models automatically constructed from data set (A). H, I and J only differ in the height that is extrapolated from (A) (H: 3D with a height of 4 cell layers, I: 3D with a height of 10 cell layers). In H-J model cells were omitted to reveal the sinusoidal network. The coloring of the network in H-J illustrates the predicted oxygen concentration within the sinusoids (blue = high concentration in portal field, red = low concentration near the central veins). All simulations were carried out for the whole lobe but only half of the lobe was visualized. (J) Magnified sinusoidal network within the lobe model. The sinusoids were not directly reconstructed from bright field micrographs but are based on the statistical data obtained from the corresponding three-dimensional volume datasets obtained by confocal fluorescence microscopy [32].

## Material and Methods

### Experiments

Male C57BL/6N mice, 11-14 weeks old (Charles River, Sulzfeld, Germany) were used. The mice were fed ad libitum with Ssniff R/M-H, 10 mm standard diet (Ssniff, Soest, Germany) and all experiments were approved by the local authorities. After the specified period of time the mice were killed by neck dislocation. The abdominal cavity was immediately opened and the whole liver was carefully excised without damaging the liver capsule. Then the tissue was separated into three different parts. One part of about 5 mm$^3$ in size was frozen in liquid nitrogen and then stored at -80°C for later RNA isolation. The two larger parts of about 1 cm$^3$ in size were used for immunohistochemical analysis. For the preparation of vibratome slices one of them was collected in 4% paraformaldehyde (Sigma, Munich, Germany) and penetrated for 48h at room temperature and then stored in PBS at 4°C until further use. The latter part of the liver was embedded into paraffin. For this purpose it was transferred to paraffin embedding cassettes (Carl-Roth, Karlsruhe, Germany) and stored in 4% paraformaldehyde for 48h at 4°C. Formalin-fixed liver tissue was washed in PBS for 48h, dehydrated through an ethanol gradient (four times 5 min in 70%, 90% and 95% ethanol, respectively, followed by three times in 100% ethanol). Subsequently, tissue specimens were incubated four times in xylene (Carl-Roth, Karlsruhe, Germany) and incubated overnight in xylene/paraffin (1:1) at 60°C. Afterwards, tissue specimens were incubated twice in 60°C paraffin for 3h, followed by embedding in paraffin.

Slices of 5 µm were prepared using a microtome (Microm, Walldorf, Germany) mounted onto SuperfrostPlus slides, and heated for 20 min at 60°C. Sections were then deparaffinized by five times washing in Rotihistol (Carl-Roth, Karlsruhe, Germany) for 5 min each, followed by hydration through a descending ethanol gradient (100%, 95%, 90%, and 70% ethanol for 5 min each) and 5 min in PBS. During the next step the sections were boiled twice in a microwave oven for 7min in 0.01M citrate buffer (Carl-Roth, Karlsruhe, Germany; pH 6.0). Endogenous peroxidase was blocked by 30 min incubation in a solution of 7.5% H2O2 in methanol at room temperature. All further incubations were performed in a humidified chamber. Unspecific binding sites were blocked by 3% BSA / 0.1% TweenR20 using 100µL per section. Subsequently, endogenous biotin and avidin were blocked using a commercially available kit (Avidin-/Biotin-Blocking-Kit, Vector Lab., Burlingame, USA) according to the manufacturer's instructions. Leaving out a washing step the blocking solution was dripped off carefully and the primary antibodies (rat-anti-BrdU, Serotec, Dusseldorf, Germany; 1:25 diluted) were incubated on the tissue section for 1h at room temperature. Before proceeding with the next incubation step the slides were washed three times 5min in PBS. Biotinylated secondary antibodies (Dianova, Hamburg, Germany; raised in goat, 1:250 diluted) were chosen to detect the primary antibodies. After 1h incubation at room temperature the slices were washed again for three times 5min in PBS. Streptavidin-horseraddish-peroxidase (Dianova, Hamburg, Germany; 1:500 diluted) was incubated on the tissue sections for 1h at room temperature. After three times 5 min washing in PBS the slices were incubated for 5 min at room temperature with DAB (Dako, Glostrup, Denmark) freshly prepared according to the manufacturer's instructions. Following this the slides were rinsed for 10 min under tap-water and then counter stained using Mayer's hemalum (Merck, Darmstadt, Germany) for 90s. Again, the slides were rinsed for 10 min under tap-water and then dehydrated using the graded ethanol series (70%, 90%, 95%, and 100% for 90s each) and four times 90s of Rotihistol. Using Entellan (Merck, Darmstadt, Germany) the slides were mounted and stored in the dark at room temperature until further analysis. Screening of DAB-stained slices was done using a conventional brightfield microscope (Olympus BX41). Images were acquired and organized using cell software (Olympus).

Image analysis

At the liver lobule level, we analyzed confocal volume data sets of high resolution (2048 * 2048 pixels in xy-plane) and 10x-20x magnification such that more than one individual lobule could be studied. At the lobe level, stained whole slide scans were analyzed (Fig.1A). In order to quantify the microarchitecture of groups of lobules, we used an improved variant of the image processing and analysis chain that was established in [32] (for further details also refer to the extensive supplemental information in the supplement of [32]. For example, we now used contrast limited adaptive histogram equalization (CLAHE) as elaborated in [54] to more efficiently increase and equalize the contrast in the volume datasets without too strong amplification of noise (Fig.1B, red rectangle). By analyzing groups of lobules, we were for example able to obtain a better understanding of the architecture of the sinusoidal blood vessels in the portal field. In addition to the quantification of the three-dimensional architecture of groups of lobules, we also obtained information on the lobe level by additionally analyzing whole slide scans of mouse liver lobes. These data allowed us to (1) quantify the proliferation during regeneration after partial hepatectomy and (2) automatically construct a three-dimensional cell-based model of a whole liver lobe.

Cell size analysis

Possible changes of the hepatocyte size during the regeneration were analyzed using bright field micrographs of control mice (18h after a pseudo surgery) and of mice 6 days after PHx that were i.a. stained using DAPI. After improving the micrographs using the image processing and-analysis chain described above, we used a marker-controlled watershed segmentation algorithm as in [40] to determine the position of the cell nuclei. The centers of the cell nuclei were then interpreted as Voronoi sites. The corresponding Voronoi diagram (Fig.3A) gives a good approximation of the cell area per nucleus in the cutting plane of the micrograph. Despite this information cannot be used to determine the exact cell volume or cell shape of individual cells in 3D, it can, however, be used as a first estimate for the cell size distributions for different time points during PHx and detect possible size changes. Knowing the relation between cell area in 2D sections and 3D reconstructions from confocal micrographs [32], an approximation for the volume per nucleus was inferred from the measured cell area. A more detailed analysis would have required here membrane and nucleus staining to separate between mono- and bi-nucleated cells. In the following we refer to the area per nucleus as "cell area" and to the volume per nucleus as "cell volume", which is strictly correct in case of mono-nucleated cells.

Lobule size analysis

Using a similar approach as for the cell size analysis, we studied possible changes in lobule size during the regeneration process. We analyzed GS stained bright field micrographs by a partially manual procedure. While the position of the central vein of each lobule could be automatically determined by an OTSU threshold segmentation of the pericentral GS staining (as described above), the complex shape of some lobules in the two-dimensional cutting plane required a manual assessment of the localization of the corresponding portal veins. The size of the lobules was then automatically calculated from these locations by obtaining the area within the convex hull [9] of the obtained points. Similar to the previous section, the exact calculation of the three-dimensional shape and volume of individual lobules is not possible based on two-dimensional micrographs but the distribution of lobule areas in the cutting plane can be used to robustly estimate the growth of the lobules in 3D. We studied the area of the liver lobules during the first 4 days after PHx (Fig.3C) and found that their size increases significantly. An immediate implication of this finding, together with the finding that the cell area per nucleus remains unchanged, is that it is proliferation and not increase in volume per nucleus (which in the case of only mononuclear cells would correspond to the cell volume) that leads to the growth of the liver remnant after PHx. Note, that if the first wave of proliferations were by bi-nucleated cells that divide into two mononucleated cells by cytoplasmic division, the lobule volume would not significantly increase. Thus proliferation (growth and division) is necessary to increase the lobule volume.

### Distribution of proliferation

In a next step, we quantified proliferation during the regeneration process to further parameterize our model. We analyzed whole slide scans of mouse lobes that were stained with BrdU / PCNA and GS. After identifying the hepatocyte nuclei within the images using a marker-controlled watershed segmentation algorithm as in [40], we used an OTSU threshold to decide whether a cell nucleus is BrdU (or PCNA) positive or not. By combining this data with information about the localization of the central veins in the lobe (based on the GS staining), we were able to quantify proliferation (A) within the lobe e.g. in relation to the distance of proliferating cells to the Glisson capsule and (B) within the lobules e.g. in relation to the distances to the central vein and the portal field. Interestingly, for mouse we found a homogeneously distributed proliferation both within the lobe and within the individual lobules.

## Model description

### Model at the liver lobule level

At the lobular level, we closely follow the model description in [31], with some modifications based on [51].

#### (A-1) Hepatocyte cell shape and physical forces

Hepatocytes in 3D culture adopt an almost perfect spherical shape (supplement in [32]). In *in-vivo* tissues visualized by confocal micrographs, hepatocytes adopt shapes reminiscent of densely packed deformed spheres [32]. Therefore, we assumed that hepatocytes can be modelled as homogeneous, isotropic elastic, adhesive, intrinsically spherical, and moderately deformable objects capable of migration, growth, division and death. Hepatocyte-hepatocyte and hepatocyte-blood vessel interaction forces were mimicked by the Johnson-Kendall-Roberts (JKR) [20], which could be shown by [14] to apply to living cells if compression and pulling of one cell with respect to the other cell is sufficiently fast. It shows a hysteresis behavior depending on whether the two cells approach each other or are pulled apart. For example, cohering cells when pulled apart still cohere beyond the distance at which they came into contact when they were approached. However, upon strong compression the JKR force underestimates the resistance of cells [52]. This can be balanced by phenomenologically upregulation of the Young modulus with decreasing cell-cell distance [51]. We considered simulations with and without this correction.

Hepatocytes are polar, the distribution of their cell adhesion molecules is not isotropic. We represented hepatocyte polarity by assuming that the contacts are constrained to certain regions of the hepatocyte surface. As a consequence, the force depends on the overlap of the cell surface regions where adhesive molecules are located in. In case the contact area of any of two cells in contact do not contain adhesion molecules the cohesion force is zero.

Mathematically, this was expressed as follows:

The JKR-force $F_{ij}^{JKR} = |F_{ij}^{JKR}(d_{ij})|$ where $d_{ij}$ is the distance between the centers of two interacting spheres i and j that was calculated from two implicit equations (Fig.2A):

$$\delta_{ij} = \frac{a_{ij}^2}{\tilde{R}_{ij}} - \sqrt{\frac{2\pi\hat{\gamma}_{ij}a_{ij}}{\tilde{E}_{ij}}}$$

$$a_{ij}^3 = \frac{3\tilde{R}_{ij}}{4\tilde{E}_{ij}}\left[F_{ij}^{JKR} + 3\pi\hat{\gamma}_{ij}\tilde{R}_{ij} + \sqrt{6\pi\hat{\gamma}_{ij}\tilde{R}_{ij}F_{ij}^{JKR} + \left(3\pi\hat{\gamma}_{ij}\tilde{R}_{ij}\right)^2}\right]$$

where $a_{ij}$ is the contact radius. The effective radius $\tilde{R}_{ij}$ in this equation is defined by $\tilde{R}_{ij}^{-1} = R_i^{-1} + R_j^{-1}$, where $R_i$ is the radius of cell *i*. $d_{ij} = R_i + R_j - \delta_{ij}$ is the distance between the centers of model cell *i* and cell *j*, where $\delta_{ij} = \delta_i + \delta_j$ is the sum of the deformations of each cell (upon compression it is the overlap of the two spheres) along the axis linking the centers of these cells. $\tilde{E}_{ij}$ is the composite Young modulus defined by $\tilde{E}_{ij}^{-1} = (1-\nu_i^2)E_i^{-1} + (1-\nu_j^2)E_j^{-1}$. $\nu_i$ is the Poisson ratio of cell i. We approximated $\hat{\gamma}_{ij} \approx \rho_m^{ij}W_s$ where $\rho_m^{ij}$ is the density of surface adhesion molecules acting in the contact area and $W_s$ is the energy of a single bond. The second equation could not be solved explicitly for $F_{ij}^{JKR}(d_{ij})$ if $\hat{\gamma} > 0$. It was solved first to obtain $a_{ij}(F_{ij}^{JKR})$. The value of $a_{ij}$ was then inserted into the first equation to give $\delta_{ij}(a_{ij})$, and via $d_{ij} = R_i + R_j - \delta$, $d_{ij}(a_{ij})$. $F_{ij}^{JKR}(d_{ij})$ could be obtained by plotting $F_{ij}^{JKR}(d_{ij})$ vs. $d_{ij}$. Different from previous communications we here also studied the effects of cell compression forces upon large compression, which we approximated by choosing $E_i$ as a function that increases with decreasing distance $d_{ij}$ in the equation for $\tilde{E}_{ij}$, $E_i \rightarrow E_i(1 + \alpha\delta_{ij}^4)$. Such a function captures the observations made in simulations within a computational high-resolution cell model to correct the JKR-force at high volume compressions [51]. This was necessary, as the JKR-force is based on pairwise interactions between cells, which remains a reasonable approximation for more than two interaction cells only if volume compressions remain moderate.

The effect of polarity was modelled by replacing the membrane density of adhesion molecules $\rho_m \rightarrow \rho_m A_{ij}^{adh}(\psi_{ij})/A_{ij}$, in which case only adhesion is downscaled. Here, $A_{ij}^{adh}(\psi_{ij})$ is the area of the overlapping regions that were able to form the adhesive contact within the contact area $A_{ij} \approx \pi a_{ij} \geq A_{ij}^{adh}(\psi_{ij})$. This approximation results in a reduced adhesion force if the overlap area of the membrane regions of neighboring cells carrying the adhesion molecules is smaller than the physical contact area.

In general the density of adhesion molecules on the surface of the two interacting cells differs [43, 44], so that $\rho_m^{ij}$ has to be calculated from the density of cell adhesion molecules on the surface of each individual cell (or, more general, of a cell *i* and its interaction object X). We here assumed for simplicity that all surface adhesion molecules in the contact region of a cell and its interacting object (e.g. another cell or sinusoid) were saturated. In this case the density of formed bonds behaves approximately as $\rho_m^{iX} \propto min(\rho_i, \rho_X)$. Here $\rho_i$ is the density of surface adhesion molecules of cell *i*, $\rho_X$ the density of surface adhesion molecules of object X. We further assumed that the density of adhesion molecules in the cell surface was the same for each cell. However, for the simulations of regeneration the assumptions of saturated bonds in contact zone and the same density of adhesion molecules are not critical, as the cells are compressed due to proliferation so cell-cell contacts not under tensile stress.

## (A-2) Equation of motion for cells

Migration of hepatocytes had been calculated using one equation of motion for each hepatocyte. An equation of motion permits to calculate the change of position of an object (here a hepatocyte) with time. It is obtained by denoting all forces acting on the object, including active force contributions as for example a contribution due to cell micro-motility.

Knowing the velocity $\underline{v}$ and the current position $\underline{r}$ permits to calculate the new position of the object from $d\underline{r}/dt = \underline{v}$, emerging from solving force balance.

Mathematically, the equation of motion for the cell i read:

$$\underbrace{m_i \frac{d\underline{v}_i}{dt}}_{\text{inertia}} + \underbrace{\underline{\underline{\varsigma}}_{iECM}^{CECM} \underline{v}_i(t)}_{\substack{\text{cell-ECM}\\\text{friction}}} = \underbrace{\sum_{jNNi} \underline{\underline{\varsigma}}_{ij}^{CC}(\underline{v}_j(t) - \underline{v}_i(t))}_{\substack{\text{cell-cell}\\\text{friction}}} + \underbrace{\sum_{jNNi} \underline{F}_{ij}^{CC}}_{\substack{\text{cell-cell}\\\text{adhesion \&}\\\text{repulsion}}} + \underbrace{\sum_{i} \underline{F}_{iECM}}_{\substack{\text{cell-}\\\text{substrate}\\\text{adhesion \&}\\\text{repulsion}}}$$

$$+ \underbrace{\sum_{jNNi} \underline{\underline{\varsigma}}_{ij}^{CS}(\underline{w}_j(t) - \underline{v}_i(t))}_{\substack{\text{cell-sinusoid}\\\text{friction}}} + \underbrace{\sum_{jNNi} \underline{F}_{ij}^{CS}}_{\substack{\text{cell-sinusoid}\\\text{adhesion \&}\\\text{repulsion}}} + \underbrace{\sum_{i} \underline{F}_i^{active,C}}_{\text{micro-motility}}$$

(Eqn. 1)

$\underline{v}_i(t)$ is the velocity of hepatocyte *i*. In the first sum, *j* denotes all neighbor cells of cell *i*, in the second sum, *j* denotes all sinusoidal elements interacting with cell *i*. Within tissues the friction between cells and the extracellular matrix components, and between cells and the sinusoids is large such that the inertia term, the first term in equation (1), can be neglected and be set to zero. $\underline{\underline{\varsigma}}_{iX}^{k}$ denotes the friction tensor (here a 3x3 matrix) describing the friction of hepatocytes *i* and j (for X=j, CC), or cells *i* and sinusoids (for X=j, k=CS), or hepatocytes and ECM (for X=ECM, k=CECM). The friction tensor may be decomposed into a perpendicular and parallel component: $\underline{\underline{\varsigma}}_{iX}^{k} = \gamma_{\perp}^{k,iX}(\underline{u}_{iX} \otimes \underline{u}_{iX}) + \gamma_{\parallel}^{k,iX}(\underline{\underline{I}} - \underline{u}_{iX} \otimes \underline{u}_{iX})$. Here, $\underline{u}_{iX}=(\underline{r}_X-\underline{r}_i)/|\underline{r}_X-\underline{r}_i|$ with $\underline{r}_i$ denoting the position of cell *i*. "$\otimes$" denotes the dyadic product. $\underline{F}_{iX}$ denotes the JKR-force between cells *i* and *j* (for X=j) as well as between cell *i* and substrate (for *X=s* enumerating sinusoidal elements). $\underline{\underline{I}}$ is the unity matrix (here a 3x3 matrix with "1" on the diagonal and "0" on the off-diagonals). $\gamma_\perp^{k,iX}$, $\gamma_\parallel^{k,iX}$ are the perpendicular and parallel friction coefficients, respectively. This becomes more apparent when multiplying the friction tensor by the difference in velocity between cell *i* and object *X*, $\Delta \underline{v}_{iX} = \underline{v}_X - \underline{v}_i$,

$$\underline{\underline{\varsigma}}_{iX}^k \Delta \underline{v}_i = \gamma_\perp^{k,iX}(\underline{u}_{iX} \otimes \underline{u}_{iX})\Delta \underline{v}_{ij} + \gamma_\parallel^{k,iX}(\underline{\underline{I}} - \underline{u}_{iX} \otimes \underline{u}_{iX})\Delta \underline{v}_{iX}$$
$$= \gamma_\perp^{k,iX} \underline{u}_{iX}(\underline{u}_{iX}\Delta \underline{v}_{iX}) + \gamma_\parallel^{k,iX}(\underline{\underline{I}}\Delta \underline{v}_{iX} - \underline{u}_{iX}(\underline{u}_{iX}\Delta \underline{v}_{iX}))$$

The first term on the rhs. specifies the friction perpendicular to the direction of movement difference, the second term on the rhs. the tangential friction.

$\underline{F}_i^{active,H}$ denotes the active movement force by migration and is denoted in assumption A-3.

The model assumes $\underline{\underline{\varsigma}}_{iECM} = \gamma^{CECM} \underline{\underline{I}}$ i.e., isotropic friction with the extracellular matrix in the space of Disse.

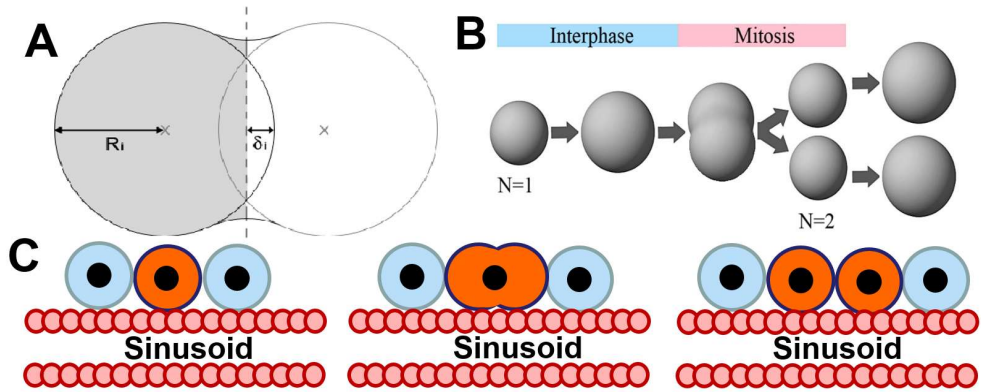

**Figure 2**: Model components. A) shows a sketch of two interacting cells for the definition of the indention δ and cell radius $R_i$. B) shows the implementation of cell growth in interface by radius increase until cell volume doubled, and division by splitting. Dividing cells align along the closest sinusoid, a mechanism we had named "HSA" [32].

Generally, the perpendicular and parallel friction coefficients, $\gamma_\perp^{k,iX}$, $\gamma_\parallel^{k,iX}$ is different for each type of interaction (k=CC, CS, CECM), and depend on the mechanisms of friction. For example, for adhesion controlled cell-cell friction one might expect $\gamma_\parallel^{k,ij} = A_{ij}^{adh}\rho_{ij}^m \zeta_\parallel^k$ with k=CC. I.e., friction basically depends on the shared contact area decorated with adhesive bonds, the density of adhesive bonds formed, and an unknown coefficient that characterizes the strength of friction between two cells, $\zeta_\parallel^{CC}$. The density of surface adhesion molecules and friction coefficient were lumped together by setting $\gamma_\parallel^{k,iX} = A_{ij}^{ad} \xi_\parallel^k$ with $\xi_\parallel^k = \rho_{ij}^m\zeta_\parallel^k$ and $\gamma_\perp^{k,iX} = A_{ij}^{ad} \xi_\perp^k$ with $\xi_\perp^k = \rho_{ij}^m\zeta_\perp^k$. Moreover, we chose $\xi_\perp^k = \xi_\parallel^k \equiv \xi^k$ with k=CC, CS, CECM.

### (A-3) Cell migration

Cells migrate actively. In the absence of morphogen gradients active cell micro-motility was assumed to be random and isotropic (uniform).
In the presence of a morphogen gradient, the cells were assumed to actively move up the gradient by chemotaxis hence in that case migration was directed in accordance with the finding in [32].

In this work we tested alternatively the hypothesis that cells during migration move towards the Glisson capsule enclosing the liver lobe.

Formally, active cell movement (migration) was mimicked by $\underline{F}_i^{active,H} = \chi \nabla c + \sqrt{2D\gamma^2}\, \underline{\eta}_i(t)$.

$\underline{\eta}_i(t)$ denotes a Gaussian-distributed random variable with average $\langle \underline{\eta}_i(t) \rangle = 0$ and autocorrelation $\langle \eta_{mi}(t')\eta_{nj}(t) \rangle = \delta_{ij}\delta(t'-t)$ (m, n = x, y, z denote the coordinate direction; i, j are the hepatocyte indices). Here, $\langle \underline{X} \rangle$ denotes the expectation value obtained by averaging the random variable $\underline{X}$ over many of its realizations. As each component of $\underline{\eta}$ is Gaussian distributed, each realization is sampled from a Gaussian distribution. $D$ is the cell diffusion constant and assumed to be a scalar, $\chi$ is the chemotaxis coefficient, $c(\underline{r},t)$ the morphogen concentration secreted by the cells dying from drug damage or from the Glisson capsule as one model hypothesis we tested. The same result can be obtained if a cell at an interface to the Glisson capsule or a necrotic zone moves in a way that it escapes regions with high cell density [32]. In [32] also a pressure-based migration mechanism had been established, that for the results presented in this work yielded the same outcome and is therefore not detailed in this work.

(A-4) Cell orientation changes

Cell orientation changes can be modeled by an optimization process based on the energy change occurring if the cell orientation changes [21], or an equation for the angular momentum [20]. The energy can be calculated from the forces by integration of the energy over the path. The energy-based method is much easier to evaluate and leads to equivalent results, which is why we used it here. Fundamentally, orientation changes were assumed to be driven by energy minimization for which we used the Metropolis algorithm [20]. In the Metropolis algorithm a trial step (here: a small rotation) was performed, and subsequently it was evaluated whether this step was accepted, or rejected (in which case the step is taken back). The change of total energy of the whole cell configuration was used to evaluate the step. As the orientation change of a hepatocyte only affects the next and maybe next-next neighbors, only those neighbors needed to be considered.

In mathematical terms, to calculate the orientation change of a cell, within each time interval Δt for each hepatocyte a rotation trial around three space-fixed axes by angles $\delta\theta_i$ with i=1, 2, 3, $\delta\omega_i \in [0, \delta\omega_{max}]$, with $\delta\omega_{max} \ll \pi/2$ was performed, using the algorithm of Barker and Watts (explained in [21]).

The energy was calculated by integration of the equation $\underline{F}_{ij} = -\frac{\partial V_{ij}}{\partial \underline{r}_i}$ where only the JKR-force contributions were considered. The energy difference is then calculated from $\Delta V_{ij}(t) = V_{ij}(t+\Delta t) - V_{ij}(t)$, and the probability that a step was accepted was calculated using $p = \min(1, e^{-\Delta V_{ij}/F_T})$ where $F_T \approx 10^{-16}$ J is a reference energy (comparable to the $k_b T$ in fluids or gases were $k_b$ is the Boltzmann factor, $T$ the temperature).

(A-5) Cell cycle progression & division

During $G_1$, S, and $G_2$-phase (interphase) we assumed that a cell doubled its volume, and then deformed into a dumb-bell at constant volume until division (Fig. 2B).

Different from [32] we studied the effect of pressure-inhibited cell cycle progression (BGC) by assuming that a hepatocyte *i* does not re-enter the cell cycle if the pressure exerted on it overcomes a threshold value.

The decision of whether a hepatocyte of a regenerating liver lobule enters into the cell cycle was now made in two steps:

**A5(a)**: Sampling of candidate cells for cell cycle entrance. For this prior to start of the simulation a cell cycle entrance rate *k* and a fixed time step *Δt* are fixed such that $N_{max}kΔt \ll 1$ for the maximal cell population number $N=N_{max}$ that might be reached in the simulation. Then, at any time step *Δt* it is calculated whether in that time interval a cell enters the cell cycle by (1.) choosing a uniformly distributed random number $\zeta \in [0,1)$ and (2.) determining if $\zeta < NkΔt$ for the actual cell population size N. If the latter conditions apply, one cell out of the N cells is picked at random and marked as a candidate cell for cell cycle entrance. The chosen process determines a Poisson process, however, small modifications reducing noise may be chosen [21] without expecting major differences.

**A5(b)**: Permitting a candidate cell to enter the cell cycle the mechanical pressure on it was below a threshold value $p^{th}$. If condition (A5(b)) was not met, the cell did not enter the cell cycle. Absence of BGC was modelled formally by setting the threshold pressure to infinity ($p^{th} \to \infty$), presence of BGC by choosing the pressure threshold to smaller values.

Note here that pressure is closely related to volume by the cell compressibility. We expect therefore that a volume threshold would have led to the same qualitative scenarios.

Mathematically, during interphase, a cell increased its volume by increasing its radius R in small steps ΔR << R until it had doubled its initial ``intrinsic´´ volume to $V_{DIV} = 2V_{INIT}$, where $V_{INIT}$ was its volume immediately after cell division (Fig.2B). Here, the intrinsic volume $V_i$ of a model cell *i* was approximated by $V_i(R_i)=4\pi R_i^3/3$. If $V_i=V_{DIV}$ (hence $R_{DIV} \approx 1.26 R$) the model cell *i* deformed into a dumbbell at constant volume in mitosis (Fig.2B). Subsequently, it divided into two daughter cells of radius *R*. The duration *T* of the cell cycle was stochastic, sampled from a Gaussian distribution with expectation value *τ* and variance $Δτ = 2h$ additionally cropping outside the interval $T \in [\tau - Δτ, \tau + Δτ]$

Pressure was defined by the simplified measure $p_i = \sum_j \frac{F_{ij}^{CX} u_{ij}}{A_{ij}}$. Here, $\underline{F}_{ij}^{CX}$ denotes the interaction force between a cell i and object j (X denotes an object which can be a cell or a piece of the sinusoid, see A-8), $\underline{u}_{ij}$ the normal vector pointing from cell i to object j, $A_{ij}$ the interface between cell i and object j. An alternative, more sophisticated way to associate a pressure to each cell would have been using the virial stress tensor [51]. Its trace is the homeostatic pressure. A cell volume could then be associated with the stress by using the relation $dp_i/dV_i = -K_i/V_i$, whereby p$_i$ is pressure on the cell, V$_i$ the cell volume, and $K_i = E_i/[3(1-2\nu_i)]$ the compression modulus of cell i. However, the pressure values emerging from this measure and the simpler one in this paper behave proportionally [52].

(A-6) Cell orientation during division

In agreement with the findings in liver regeneration after drug induced liver damage, we assumed that hepatocytes divide along the closest sinusoid (named hepatocyte-sinusoid alignment, Fig.2C) [32].

(A-7) Sinusoids (blood micro vessels)

The model only considered sinusoids and hepatocytes, the main constituents in a liver lobule. Other cell types such as hepatic stellate cells, Kupffer cells or externally invading macrophages were neglected as these were not needed to explain the principle of regeneration of liver mass and architecture after drug-induced liver damage.

Sinusoids were mimicked as a chain of spheres with a radius equivalent to that obtained by inscribing spheres into vessel segmentations within full volume data sets reconstructed from confocal laser scanning images. Neighbor spheres were pairwise linked by linear springs whereby the spring constant was chosen to reproduce a certain range of Young moduli.

In mathematical terms, each of the sinusoidal spherical elements was assumed to interact with the hepatocytes by a JKR-force ($\underline{F}_{ij} = \underline{F}(d_{ij}, \psi_{ij})$). The forces among sinusoidal elements was approximated by linear elastic springs, $\underline{F}^S_{kl} = -\frac{kl_0}{A} \times (\frac{l^S_{kl}}{l_0} - 1)\underline{u}_{kl}$ with $k, l$ being spheres on the chain connected by a spring, $A = \pi r^S_{kl}$ is the sinusoid element intersection area with $r^S_{kl}$ being the radius of the sinusoid element connecting points $k$ and $l$ (in [32] we used a constant sinusoid radius, see Tab.1). $l_0$ is the spring rest length, $l^S_{kl}$ the actual length. The spring and geometrical parameters can be related to the (elastic) Young modulus by setting $E^S = \frac{kl_0}{A}$. The Young modulus is one model parameter. $\underline{u}_{kl}$ is the unit vector pointing from the center of sinusoidal object $k$ to sinusoidal object $l$.

Movement of the sinusoids is modelled by an equation of motion for each of the sinusoidal spheroid elements using the same type of equation as in for the hepatocytes except for sinusoid we missed out an active motion (migration) force.

Sinusoids in the model were anchored in the central vein, and in the portal veins.

### (A-8) Reference parameters

All parameters in the model defined above have either a direct biophysical or a bio-kinetic interpretation, and in principle could be determined experimentally. Thus, the physiologically meaningful parameter range for each of the parameters could be estimated. As reference parameters we used the parameter set, for which we had found the best agreement between model simulations and experimental data in regeneration after drug-induced peri-central liver lobule damage in the mouse model [32]. This set of parameters was found by simulated parameter sensitivity analysis varying each model parameter within its physiologically meaningful range, followed by direct comparison of the model simulation outcome with experimental results. By this sensitivity analysis that could be embedded in a general model identification strategy [22] we were able to rule out model mechanisms that were insufficient in explaining the biological data, and identify the minimal model and its parameters for which the experimental findings could be quantitatively explained.

## Model at the liver lobe level

### (B-1) Lobe model construction by image processing and -analysis

The model of the liver lobe represented a larger portion of liver tissue composed of many liver lobules enclosed by an elastic capsule. Each lobule was modelled as described in the previous section. The sinusoidal network of the individual lobules was generated by a novel species-dependent extension of the vessel generator that was previously introduced in [32]. This extension was used to create an interconnected sinusoidal network for an arbitrary number of lobules thereby constituting either an entire model liver lobe (Fig.1).or a representative part of it that fully represented the network statistics of in-vivo sinusoidal networks (SI-Fig.2A/B).

We parameterized the liver lobe model by quantifying experimental liver lobe micrographs, segmenting all central veins, and portal veins or hepatic arteries using them as base points for an approximation of liver lobule shape as described in [26]. The model construction algorithm used two neighboring whole slide scans of liver lobes (Fig.1), one reference slide as the basis for the reconstruction and the other providing complementary information. The reference slide could for example have been stained for proliferating cells using PCNA (Fig.1A), while the second slide could have been stained for glutamine synthetase (GS) to permit distinguishing between central and portal vessels (Fig.1B). In principle, the algorithm also works with only one single GS-stained slide, but we found that two neighboring slides, one GS-stained and other not GS-stained generally produce better results due to the easier cell nuclei segmentation of non-GS-stained hepatocytes. In a first step the whole slide scans were preprocessed using the improved image processing chain to prepare the following analysis e.g. by improving the contrast of the scans using CLAHE (Fig.1B, red rectangle). In the next step, the two slides were rigidly registered thus merging the information from the GS staining into the first base slide (Fig.1C, green).

Utilizing this information, we determine whether a larger vessel should be considered a central vein or a part of a portal triad by the area of GS-positive tissue in its vicinity. We found that the complex three-dimensional architecture of the liver vascular system may lead to configurations where such decision was very hard if not impossible to take with certainty on the basis of two-dimensional images. A solution would be to use three-dimensional imaging for the entire lobe, which was not available. Alternatively, in these cases, manual adjustments of experts would then be required. In most situations, however, central veins can be robustly and automatically distinguished from portal veins and arteries.

In a next step, we quantified the cell nuclei using a marker-controlled watershed segmentation algorithm as in [40]. We calculated the positions of the corresponding hepatocytes as centers of the Voronoi cells of the Voronoi diagram that could be constructed using the centers of the segmented cell nuclei as Voronoi sites.

Since a lobe model integrating an entire lobe volume would be too computation-time intense, we considered a slice of a lobe (Fig. 1D) and varied the thickness in the simulations to test at which thickness the simulation results became independent of the slice thickness. We used a linear extrapolation of 2D information to 3D and studied thicknesses of the model liver lobe of one (Fig. 1D), two, four, and 10 hepatocyte diameters (SI-Fig.3A).

(B-2) Glisson capsule

Additionally, the liver lobe model represents the Glisson capsule by a system of springs whereby the spring constant was varied to represent different resistances effectively mimicking an elastic layer around the model lobe (Fig. 4C). An intensity-based OTSU segmentation identified the shape of the lobe which was assumed to represent the position of the Glisson capsule (Fig.1C, red) that generally encapsulates the whole lobe: In a mathematical model on the scale of a lobe, the mechanical impact of the Glisson capsule (E(Capsule)≈400 kPa, [12]) may be significant and thus must be taken into account. Using the obtained information on (1) the shape of the lobe (B-2) and its capsule, (2) the position, shape and type of the larger vessels within the lobe, and (3) the position of the cells within the lobe (B-1), we are able to automatically construct a corresponding cell-based model that reflects all of these aspects (Fig.1D-J).

| Parameter | Mouse Source | Mouse Value ± Standard deviation | Pig Source | Pig Value ± Standard deviation |
|---|---|---|---|---|
| Number of analyzed datasets | - | 26 | - | 4 |
| **Lobule** | | | | |
| Confocal scanning depth | Confocal metadata | 95 ± 57 µm | Confocal metadata | 60.21 ± 9.11 µm |
| Lobule height in the model | 10 Cell layers | 250 ± 0 µm | 5 Cell layers | 90 ± 0 µm |
| Lobule area (2D slice) | Bright field microscopy | 0.21 ± 0.05 mm² | Bright field whole slides (Sirius Red) | 1.66 ± 0.84 mm² |
| Lobule radius in model (2D slice) | $R = \sqrt{\dfrac{2 \cdot A}{3 \cdot \sqrt{3}}}$  $A$... lobule area, assuming a regular hexagon | 284.3 ± 56.9 µm (12.2 ± 2.4 hepatocytes) | Calculations as in mouse | 799.33 ± 260.7 µm (43.0 ± 14.6 hepatocytes) |
| Area of necrotic lesion before regeneration | Image analysis | 0.073 ± 0.011 mm² | -[1] | - |
| Radius of necrotic lesion before regeneration | $R_{nec} = \sqrt{\dfrac{A_{nec}}{\pi}}$  (assuming a circular necrotic lesion) | 149 ± 22 µm (6.4 ± 1.0 hepatocytes) | -[1] | - |
| **Sinusoids** | | | | |
| Radius of sinusoid vessels | Volume analysis | 4.75 ± 2.25 µm | Volume analysis | 8.85 ± 3.11 µm |
| Orthogonal minimal vessel distance | Volume analysis | 16.45 ± 4.22 µm | Volume analysis | 25.7 ± 8.28 µm |
| Non-branched segment length | Volume analysis | 43.1 ± 18.9 µm | -[2] | - |
| Mean branching angles | Volume analysis | 32.5° ± 11.2° | -[2] | - |
| Vessel volume in lobule | Volume analysis | 7.4 ± 1.1% | Volume analysis | 11.1 ± 2.6% |
| **Hepatocytes** | | | | |
| Hepatocyte volume | Volume analysis | $1.2653 \cdot 10^{-5} \pm 3.915 \cdot 10^{-6}$ mm³ | Volume analysis | $0.6838 \cdot 10^{-5} \pm 8.133 \cdot 10^{-6}$ mm³ |
| Hepatocyte diameter | Volume analysis | 23.3 ± 3.1 µm | Volume analysis | 18.6 ± 4.7 µm |
| Hepatocyte density | Image analysis | 1889 ± 341 cells/mm² | Volume analysis | 2631 ± 397 cells/mm² |
| Next neighbor distance | Volume analysis | 21.6 ± 13.1 µm | Volume analysis | 17.4 ± 11.5 µm |
| Diameter of hepatocyte nucleus | Image analysis | 9.3 ± 4.4 µm | -[1] | - |
| **Central vein** | | | | |
| Length in Volume | Volume analysis | 107 ± 69 µm | -[1] | - |
| Radius | Volume analysis | 41.2 ± 32.1 µm | Manual analysis | 45.3 ± 52.1 µm |
| Inclination to viewing plane | Volume analysis | 6.6° ± 4.1° | -[1] | - |
| **Capsule** | | | | |
| Capsule thickness | Manual analysis (IfaDO, 5 datasets in 2D) | 10.75 ± 1.14 µm | Manual analysis (IfaDO, 5 datasets in 2D) | 24.96 ± 9.06 µm |

[1] Not relevant for current simulations
[2] Not yet possible with current datasets (too limited sinusoid tracking)

**Table 1**: Lobule parameters for mouse and pig

# Results

## Preparation of experimental data by image processing and -analysis

In conceptual analogy with the analysis of the regeneration process after drug induced damage [22, 32] we defined several process parameters to characterize the regeneration process to set up and parameterize a predictive mathematical model of liver regeneration after partial hepatectomy (PHx). The process parameters chosen for regeneration after PHx were (i) the cell size (PPi), (ii) the lobule size (PPii), (iii) the lobe size (PPiii), and (iv) the proliferation pattern (PPiv). This information was extracted from histological tissue slides of mouse liver tissue. The data was then used to parameterize the model.

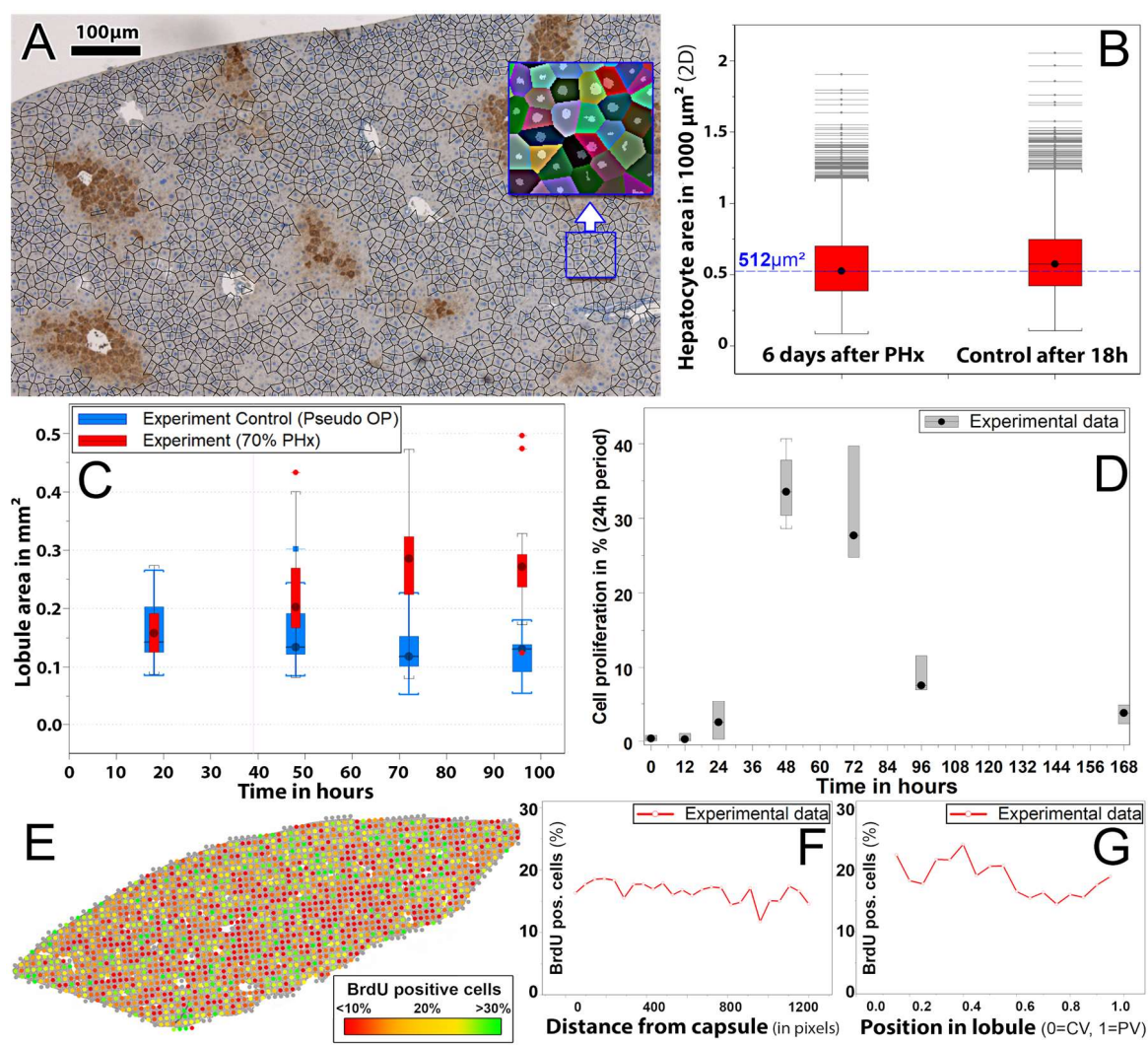

**Figure 3**: Experimental parameters in regenerating liver tissue of mice. (A) Representative DAPI and GS stained bright field micrograph. (B) A number of similar micrographs were used to study cell size distributions during regeneration after PHx. The blue line represents the model = 512µm²)). (C) Lobule size in control mice (blue) and mice that underwent PHx (red). The area of the lobules increased during regeneration by a factor of approximately two. Accordingly, the volume increased by about a factor of 3 as expected after 2/3 hepatectomy. (D) Kinetics of proliferation. (E) Bright field micrograph with overlaid proliferation quantification averaged for square-shaped regions of 100 x 100 µm (one dot per region) within the lobe micrograph. The color of the dots represents the average fraction of proliferating cells within the corresponding region (green: >30% proliferating cells, red: <10% proliferating cells). (F, G) Distribution of BrdU-positive cells as measure for proliferation within (F) the lobe, (G) the individual lobules.

### Hepatocyte size

First, possible changes of the hepatocyte size during regeneration were studied (Fig.3A). We analyzed tissue consisting of more than $10^5$ cells and observed that the average hepatocyte size and size distribution underwent no major changes in the early phase (until 6 days) of the regeneration (Fig. 3B). In this phase the remnant liver lobes grew until the original liver mass was almost recovered. This suggests that the increase of the volume of the liver lobe occurred mainly by proliferation of hepatocytes that divide and grow until they reach their original size, and not by size increase of remaining hepatocytes. Hence, we assumed in our model that the average size of hepatocytes does not change during the regeneration process.

### Lobule size

For the determination of possible changes in lobule size during the regeneration process first GS stained bright field micrographs were analyzed to determine the positions of central and portal veins. The size of the lobules was then automatically calculated from these locations. The exact calculation of the three-dimensional shape and volume of individual lobules was not possible on the basis of two-dimensional micrographs but the distribution of lobule areas in the cutting plane could be used to robustly estimate the growth of the lobules in 3D. The size of the liver lobules during the first 4 days after PHx increased significantly, whereby the difference in lobule area at days 3 and 4 was negligible. The experimentally obtained lobule area increase explained almost the entire regeneration of the liver mass after PHx i.e., after reducing the liver mass to ~33% of its original value (Fig.3C); in case the lobe would grow isotopically, such an area growth until 96h would account for 82-94% of mass recovery. An implication of this observation, together with the finding that the cell size remained unchanged, was that it was proliferation and not the volume increase of existing hepatocytes that led to the growth of the liver remnant after PHx.

### Kinetics of proliferation

As a further process parameter, the percentage of proliferating cells in the entire lobe was experimentally determined at 12, 24, 48, 72, 96 and 168 hours after PHx (Fig.3D). The strongest proliferation occurred at day 2 and 3 followed by a decrease. This temporal proliferation pattern was reminiscent of the cell proliferation pattern after CCl4-induced pericentral damage, which also increases significantly at day 2 and drops at day 4 [32].

### Spatial distribution of proliferation

In a next step, the spatio-temporal distribution of proliferation during the regeneration process was quantified to further parameterize the model. For this, whole slide scans of mouse lobes stained with BrdU or PCNA and GS were analyzed. After identifying the hepatocyte nuclei, an intensity-based threshold was used to decide whether a cell nucleus is BrdU (or PCNA) positive or not (both BrdU and PCNA were used to stain proliferating cells). By combining this data with information about the localization of the central veins in the lobe (based on the GS staining) we were able to quantify the spatial proliferation pattern (A) within the lobe in relation to the closest distance of proliferating cells to the Glisson capsule, and (B) within the lobules in relation to the distance to the closest central vein and portal field. A homogeneously distributed proliferation both within the lobe and within the individual lobules was obtained (Fig.3E-G).

In summary, the data of regeneration after ⅔ PHx suggested that the liver recovers most of its mass within 4 days by mainly increasing each lobule by cell proliferation, as the average hepatocyte size and the size distribution does not change. Proliferation of hepatocytes peaked at days 2 and 3 after PHx, and was approximately homogeneously distributed over lobes and lobules.

## Simulation of the regenerating liver lobe

The next step was to study whether the experimental observations could consistently be explained in a spatial-temporal mathematical model that resolved cell scale. A biophysical cell-based model was applied representing each hepatocyte as individual modeling unit parameterized by measurable biomechanical and biokinetic parameters. Furthermore, the model included sinusoids, and central as well as portal vessels. Each model hepatocyte was able to move, grow and divide. Movement of a hepatocyte in the model was a consequence of forces exerted on the hepatocyte, by other hepatocytes, by the sinusoids, and by extracellular material, as well as of hepatocyte micro-motility represented as an active force in the equation simulating hepatocyte movement. Sinusoids were mimicked as elastic chains anchored in the central and portal veins. Other cell types have not been explicitly represented. At lobule scale, the model largely corresponded to the experimentally validated model of the regenerating liver lobule after administration of the hepatotoxic compound $CCl_4$ (Hoehme et. al., 2010) (SI-Fig.2). As the PHx experiments have been executed in the same animal model as previously the experiments on drug-induced damage after CCl4, we varied only those cell- and sinusoid model parameters at the lobule scale, that were observed or expected to change, as in particular the spatial-temporal cell cycle progression pattern, the size of the sinusoidal network, and the lobule size. These parameters were varied within physiological meaningful ranges. At the lobe scale, the model represented a longitudinal section through an entire liver lobe with all its lobules, and included the Glisson capsule (Fig. 4C). The thickness of the section was varied in the simulations to identify the minimum lobe thickness at which the simulation results no longer depended on that parameter (SI-Fig.3). We assumed that proliferation is stopped as soon as the original mass of the liver has been restored (at a lobule area of ~0.3mm$^2$). For the control curve, mice underwent a pseudo surgery whereby no part of the liver was removed (Fig. 3C). In the model the proliferation rate in homeostasis was chosen to 0.001 per day according to [41].

## Simulation of liver regeneration in absence of BGC

As liver lobule regeneration after an overdose of $CCl_4$, which induces a pericentral necrosis [32], could be explained without assuming BGC, we first studied whether the data in section 3.1 on liver lobe regeneration after PHx could be explained in the absence of BGC. Simulation of the volume of an entire liver lobe was not feasible in a reasonable time. Therefore, we studied a slice of a certain thickness of an entire liver lobe using the whole slide bright field scan (Fig.1A) as a starting scenario. Central and portal vessels were identified and the internal lobule structure was obtained by a published algorithm to generate statistically representative liver lobules [32] (Fig.1J, SI-Fig.2A/B). The hepatocytes were inserted in between the vessels. The remaining fit parameter was the cell cycle entrance rate, which was varied until the average lobule size in the simulation matched that of the experiment (Fig. 4A). To exclude artifacts by a too small slide thickness, we repeated the simulation procedure with different thickness values (SI-Fig.3A). For slides with a thickness corresponding to four or more cells the results of the simulation became independent of the slide thickness.

Cell cycle entrance in absence of BGC was mimicked by randomly selecting cells at a certain rate *k* for cell cycle entrance as explained in assumptions A5a, b assuming that each cell entering the candidate phase enters the cell cycle (formally setting the pressure threshold $p^{th} \to \infty$ in A5b).

However, for a cell cycle entrance rate that was large enough to reproduce the recovery of liver lobule size in absence of BGC, the cell volume decreased down to 40% of the volume in a relaxed state (Fig. 4G). This finding was in disagreement with experimental observations of unchanged average cell size (Fig 3B). An occupation of the Disse space by the hepatocytes would not be able to account for an average reduction of area per nucleus by 40% as the Disse space with a diameter of about 0.5 µm is much too small to provide sufficient volume for the proliferating cells. Compression of the simulated hepatocytes resulted from the high pressure in the interior the lobule (Fig. 4F), which built up as a consequence of hepatocyte proliferation and could not be relaxed sufficiently fast by pushing the cells towards the borders and by expanding the lobe. A consequence of compression was that the growth rate necessary to expand the lobule within the experimentally observed time period (Fig. 4A) needed to be larger than the experimentally observed growth rate (Fig. 4L) in order to compensate for the decrease in volume due to compression. Moreover, as cells in the lobe center were much more compressed than those in the periphery of the lobe, the density of proliferation events was predicted to be higher in the lobe center than the lobe periphery (Fig. 4K). Such a compression was experimentally not observed. It might be possible that endothelial cell proliferation is delayed in comparison with hepatocyte proliferation. In this case the sinusoids would have to be stretched and get narrower i.e., the diameter of the sinusoids could be reduced. In the extreme case where the blood inside the sinusoids would give no resistance, this could increase the volume available for hepatocytes up to about 13%, which is insufficient to compensate for the 60% volume reduction observed in the simulations. Hence, we concluded that cell cycle progression in regeneration after PHx must be controlled by a mechanism that inhibits the build-up of too high pressure and too high-volume compression.

Simulation of liver regeneration with simulation in presence of BGC

In a next step biomechanical control of cell cycle progression was included into the model to study if this mechanism allowed to avoid unphysiological cell compressions and consistently explain the experimental data. With BGC now only those cells entered the cell cycle that experienced a pressure *p* below the inhibitory threshold $p_{th}$ that is chosen of the order of a few hundreds of pascal (Fig. 1F/G, Fig.4H) while without BGC (formally equivalent to setting $p_{th} \to \infty$) every cell was able to enter the cell cycle independent of the pressure it experienced. Hence in presence of BGC the cell cycle entrance is controlled by two parameters, an intrinsic proliferation rate *k* (cf. A5a), and the pressure threshold $p_{th}$ (cf. A5b). *k* determines the rate at which a cell enters a "candidate phase", $p_{th}$ whether a cell in the candidate phase enters the cell cycle i.e., gets a "GO". The model does not specify the molecular origin of the processes. However, a possible origin for entering the candidate phase might be receipt of growth signals that are required but not sufficient for the cell to commit to the cell cycle, while a possible origin for the second decision process might be related to mechanotransduction as a negative regulator. The proliferating index quantifies the fraction of proliferating cells. Hence the proliferative index is controlled by both intrinsic proliferating rate and pressure threshold. Cell divisions and local re-arrangements lead to stochastic fluctuations of the local pressure. For a given pressure threshold $p_{th}$ the chance, that the local pressure on a cell in the candidate phase is smaller than the pressure threshold, increases with the number of cells in the candidate phase. This number increases with *k*. If *k=0*, no cell enters the cell cycle.

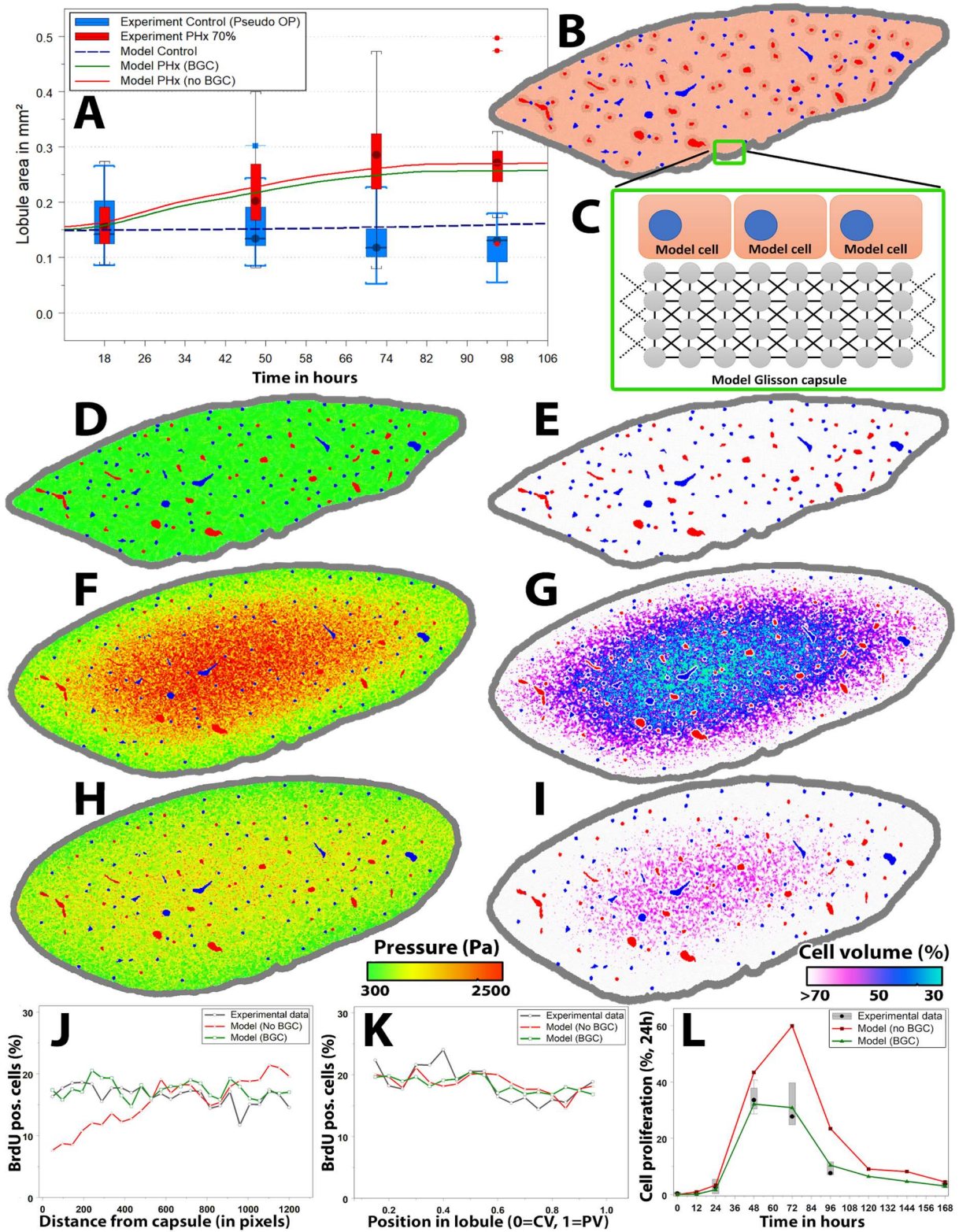

**Figure 4**: (A) Recovery of liver lobule size comparing the model simulation to experimental data. (B) Model lobe architecture at t=0 days (model initial state). (C) Illustration of the model of the Glisson capsule (D) pressure and (E) cell volume visualization at t=0. (F) Pressure and (G) cell volume visualization at t=4 days without BGC. (H) Pressure and (I) cell volume visualization at t=4 days with BGC enabled. Cell volume predictions were based on Voronoi space subdivision in the lobe model. Red = central veins, Blue = portal veins. Cell volume coloring: White: Cell volume of more than 70% of an isolated (uncompressed) cell, magenta: 50%, blue: 40%, cyan: 30% (see legend). Without BGC, the model shows unrealistically small cell volumes within the lobe. Pressure coloring: Green = Low pressure, Yellow = Intermediate, Red = High pressure. The compression in presence of BGC (H/I) is lower than in absence of BGC.

The model with BGC, was able to correctly simulate the experimentally observed growth kinetics of the lobules (Fig.4A) and the cell proliferation kinetics (Fig.4J-L) simultaneously in the same simulation. Moreover, the cell compression turned out to be significantly reduced (Fig.4I). BGC ensured that forces emerging from volume increase through cell growth and division in the interior of the lobe do not increase to unrealistically large values. Also, the cell proliferation events were independent of their distance to the Glisson capsule (Fig.4J) and from their absolute position in the lobule (Fig.4J/K), in agreement with experimental data. Moreover, the volume of the remnant liver has to increase by a factor of 3 after ⅔ hepatectomy. Assuming that the hepatocyte population size is proportional to the liver volume and an approximately equal expansion of a lobe in each coordinate direction, the hepatocyte population in a cross section of the lobe should increase by a factor of about $3^{2/3} \approx 2.08$. In our simulations this was reached after about 3 days in mouse indicating that the deviation of the lobule area after 3 days from a factor of ~2.08 might be caused by a compression of the lobules that relax only slowly.

Increasing the mitotic index within a given unit of time by increase of the probability of a cell to start proliferating (the experimentally found values is 0.5, see SI-Fig.1) within 24h (this defines the intrinsic proliferation rate) increased the lobe size (SI-Fig.1B-D), but also led to short-wavelength undulations at the Glisson capsule. This is reminiscent of a buckling instability [19]. Choosing the pressure threshold too small resulted in inhibition of cell cycle progression already at low pressure and confinement of cell proliferation to a zone close to the Glisson capsule. This is found to decrease the regeneration velocity hence the lobules at the same time point after PHx too small (SI-Fig.1 G/H). Both the intrinsic proliferation rate and the pressure threshold are model fit parameters that have to be calibrated so that the mitotic index in the simulation matches that in the experimental data at realistic cell compression.

A cell that is circumvented by many proliferating, growing neighbor cells is more likely to be under large compressive stress than a cell with no proliferating neighbor cell, and is hence unlikely to enter the cell cycle itself. Consequently, BGC should favor local arrangements where a proliferating cell has only a small number of proliferating neighbor cells. This was indeed confirmed by the simulations, where at cellular resolution the model predicted a characteristic checkerboard-pattern (Fig.5B). This markedly differed from the pattern obtained if the same total number of cells in that slice entered the cell cycle randomly with equal probability independently of their neighbors, where locally accumulations of proliferating cells were observed (Fig.5A).

In order to quantify this observation, we calculated the average fraction of proliferating cells in the vicinity of a proliferating cell $f_{PP}$ as well as the histogram of the number of proliferating cells neighboring a proliferating cell in both cases (Fig.5C). We find that BGC indeed reduces the average fraction of proliferating cells in the vicinity of a proliferating cell to $f_{PP}$ =0.2418 versus $f_{PP}$ =0.3068 in the case of cell cycle entrance by pure chance, i.e., in the absence of BGC.

To see whether the difference of ~0.06 is small or significant, we searched in the next step for the theoretically smallest and largest values of $f_{PP}$ for that mitotic index of 0.3 algorithmically, and find the algorithmic minimum at $f_{PP}$ =0.2211, and the maximum at 0.3205.

Hence the minimal and maximal values for $f_{PP}$, are close to the algorithmically found extremes with and without BGC. Consistent with this finding, we verified algorithmically that BGC maximizes the number of non-proliferating cells of a proliferating cell, kept the average fraction of proliferating cells in the vicinity of a non-proliferating cell low, and the average fraction of non-proliferating cells in the vicinity of a non-proliferating cell high (Tab.2). In addition to $f_{PP}$, the histogram depicting the number of proliferating neighbor cells of a proliferating cells were measured in the presence and absence of

BGC. It showed a peak at a lower number for BGC than for random cell cycle entrance, and that BGC inhibited too many neighboring cells of a proliferating cell to enter the cell cycle (Fig.5C).

The next question was whether one would be able to identify such a checkerboard-like pattern in BrdU-stained images, as BrdU, staining S-phase only, was used to analyze the proliferation kinetics (Fig.4D). For this purpose the BrdU-staining was simulated in the same simulations that had led to Fig. 5A/B and - similar as for Fig.5C the histograms of neighborhoods be computed. However, significant differences in the number of proliferating neighbors of a proliferating cell-histograms between random and BGC-controlled cell proliferation for BrdU-stained cells could not be found (SI-Fig. 4), likely, as the difference in absolute numbers between random and BGC-controlled cell proliferation was too small.

We used a generated pressure-limited 2D cut through the mouse lobe. N(0) = 26368. The number of proliferating cells in both cases was 8102 (=30.73%) at t = 2 days. The simulation with pressure limitation (BGC) was now perfectly matching the experimental data (e.g. Fig.4). The difference with and without BGC was now visible in numbers and population (see Fig.5C).

|  | Average fraction of proliferating cells in vicinity of a proliferating cell | Average fraction of non-proliferating cells in vicinity of a proliferating cell | Average fraction of proliferating cells in vicinity of a non-proliferating cell | Average fraction of non-proliferating cells in vicinity of a non-proliferating cell |
|---|---|---|---|---|
| Random (uniform) distribution Fig.5A/C | 0.3068 | 0.6932 | 0.3076 | 0.6923 |
| Biomechanical growth control Fig.5B/D | 0.2418 | 0.7582 | 0.3352 | 0.6649 |
| Heuristic min. | 0.2211 | 0.7789 | 0.3527 | 0.6473 |

**Table 2**: Average fraction of (non)-proliferation cells in vicinity of other (non)-proliferating cells

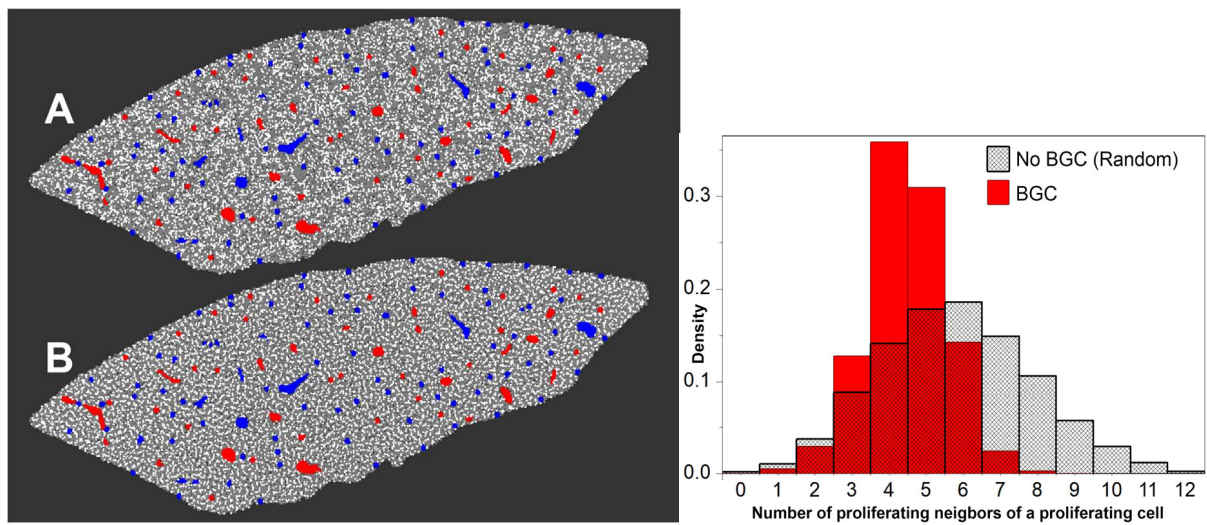

**Figure 5**: Spatial cell proliferation pattern in case of cells enter the cell cycle (A) randomly (proliferating cells in white) and (B) in the presence of BGC at time t=3d (C): Corresponding frequency histograms for the number of proliferating cells in the vicinity of a proliferating cell for BGC (pressure-based) control of cell cycle entrance and for random entrance.

In conclusion, the model of a regenerating liver lobe including BGC was able to consistently explain the experimental data and predicted a checkerboard pattern of cell proliferation that is characterized by avoiding local clusters of neighboring proliferating cells.

### Alternative mechanisms to BGC?

A further question was if possible alternative mechanisms could relax the proliferative stress. One hypothesis could be a mechanism that amplifies migration of cells towards the Glisson capsule. For example, diffusive signals entering the liver from the Glisson capsule might have made cells to migrate actively towards the Glisson capsule thereby relaxing cell compression. We tested this hypothesis assuming forces of up to ~30N as this is in the range of physiologically meaningful migration forces [51]. However, a significant relaxation permitting lobe growth without unrealistic cell compression in absence of BGC could not be found (SI-Fig.3). This indicated that for realistic force values, a directed migration was insufficient to ensure physiologic regeneration after PHx in absence of BGC.

Moreover, we tested different resistances against cell compression by smaller or larger repulsion forces as cells approach each other but also this could not account for the experimental data.

### Role of BGC in regeneration after drug-induced liver damage?

As BGC seems to be necessary to explain liver regeneration after PHx, what would be its effect on liver regeneration after drug-induced damage given in the previously created model of the liver lobule regeneration process after an overdose of CCl4 BGC had not been considered [32]. In order to study if BGC might modify the results and conclusions of that paper, all three different hypotheses of that paper were now re-simulated with BGC. It was found that BGC had no effect with the finally identified mechanisms of regeneration (denoted as model 3), while significantly impacting the mechanism underlying model 1 and moderately changing the results of model 2 (SI-Fig.2).

In conclusion a biomechanically-based cell cycle progression control mechanism (BGC) inhibits buildup of large pressure and local accumulation of proliferating cells. BGC is compatible with both regeneration of liver after drug-induced pericentral damage and after partial hepatectomy.

## Model prediction of an inhomogeneous proliferation pattern in pig

In a next step it was studied whether the model for regeneration after PHx in mouse, after reparameterization with architectural data from pig liver, could describe regeneration after PHx in pig. A pig liver lobe is much larger than a mouse liver lobe hence the displacement of cells close to the Glisson capsule necessary to recover the liver mass is significantly larger in pig than in mouse. The question was, whether this would affect the spatial regeneration pattern.

The choice of pig liver was motivated by its weight that is about the same as the liver weight of human liver, such that pig liver might be considered as a template for the translation from mouse to human. A doubling of the weight of the remnant liver during the regeneration after PHx in mice takes approximately 2 days while in humans the same doubling of weight takes about 7 days [24, 50]. At the same time the size of the individual hepatocytes is largely the same in different species [53] and the general liver architecture is largely similar. Differences are mainly in the size and number of the lobules. The reasons for the experimentally observed differences in regeneration velocity are still poorly understood.

In order to construct a predictive model for pig on the basis of the presented mouse lobe model, we re-parameterized our computational mouse model based on bright field and confocal laser scanning micrographs of pig. Image analysis of pig liver lobes stained with Sirius Red for collagen in the portal field (Fig.6A) showed that pig liver lobules are approximately 8 times larger in area compared to mouse lobules (Fig.6C) while the volume of the individual hepatocytes is largely similar. Different from mouse lobules, pig lobules are enclosed each by a capsule of connective tissue. Moreover, the microarchitecture of pig lobules might slightly differ compared to mouse lobules but our tissue samples were insufficient for a thorough statistical quantification of these differences. However, the precise lobule architecture did not seem to play a critical role in simulations of regeneration after PHx in mouse, which is why it was here assumed that the architecture is largely the same for pig than for mouse (Tab.1).

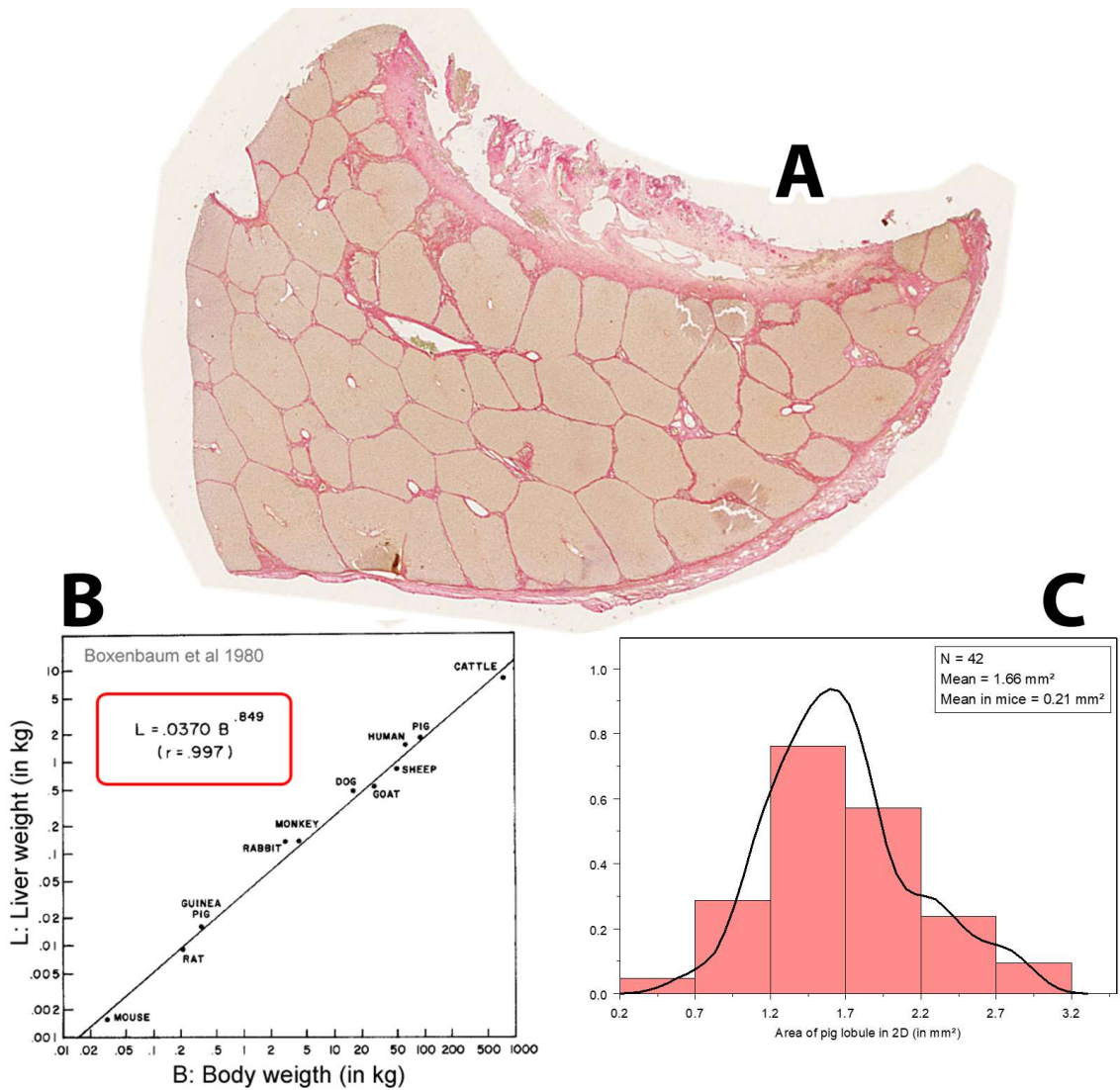

**Figure 6**: (A) Whole slide scan of a part of a pig liver lobe stained with Sirius Red for collagen in the portal field (B) Liver weight for mice and larger animals, including pig and human. (C) Area of pig liver lobules quantified from images similar to (A).

The data displayed in Tab.1 was used to construct liver lobules for pig. However, simulations of an entire pig liver lobe turned out to be not amenable to computer simulations due to the lobe size, which is why only a part of the pig lobe was simulated wherein the lobules were encapsulated by an elastic capsule (Top of Fig.7B/C). Moreover, in x-direction periodic boundary conditions were used (i.e., the cells moving out at the right border in Figs.7B/C would enter on the left border and vice-versa) and a hard, impermeable border taking into account the up-down lobe symmetry at the bottom of the simulation domain of Fig.7B/C was implemented. To verify that considering such a slice does not generate artifacts simulations with the mouse lobe model using a similar stripe geometry were performed. If the partition was chosen to be large enough (with an edge length of larger than approx. 2-3 lobules) the results were the same as for a full lobe model.

Besides the differences for the architectural parameters depicted in Tab.1 for the regenerating pig liver sample the same model parameters as for the regenerating mouse lobe have been used (e.g. the proliferation inhibition threshold was chosen for pig as for mouse in Fig.1 and Fig.4 ($p_{th}$ = 300 Pa)). This was based on the assumptions that evolutionary the cell level parameters and mechanisms should not largely differ between mouse and pig.

The result of the simulation in pig shows initially a homogeneous isotropic proliferation pattern as in mouse (Fig.7B), while after the first round of proliferation (from around day 2 on) the model predicts establishment of a proliferation gradient, which is a marked difference from regeneration in mouse (Fig.7C). In the latter phase the proliferation was predicted to be most pronounced in the vicinity of the Glisson capsule (Fig.7C).

This effect could be explained by the BGC mechanism, that inhibited cell cycle progression in case of a locally too large pressure. As pig lobes are much larger than mouse lobes, the force of growing and dividing cells in the interior of a pig lobe within the simulation was not sufficient anymore to maintain the same degree of cell cycle progression than close to the Glisson capsule. This line of argument is supported by the observation in computed parameter sensitivity analyses for the regenerating mouse lobe that proliferation of cells located in the lobe center were inhibited when the pressure threshold is too low (SI-Fig.1H). At the same time, the partial cell cycle progression inhibition in the interior of the pig lobe resulted in a decrease of the overall mitotic index compared to mouse, which might explain a slower regeneration after PHx in pig than in mouse. I.e., assuming the same pressure-threshold as in mice ($p_{th}$=300 Pa) in the much larger livers such as those of pigs or humans (Fig.6B) the volume fraction wherein the pressure would be predicted to be above the pressure threshold at which cells are able to enter the cell cycle (e.g. lower tissue region in Fig.7C) significantly increased, resulting in an overall slower regeneration after PHx in those larger species compared to smaller species. The time until the liver mass is restored can be approximately estimated from the growth of a lobe slice or part of it as follows.

In the following, this hypothesis was tested in a pilot analysis against experimental data in a single pig so far. For this purpose, PCNA stained micrographs of pig livers (whole slide scans) were studied. The staining patterns indeed suggested the possible existence of a gradient of proliferation with highest proliferation close to the Glisson capsule, as it was predicted by the model (Fig.7E-H). Moreover, the existence of such gradient can be considered as an important indicator for the validity or our hypothesis of a biomechanical growth control (BGC).

However, the current findings cannot yet be considered as final proof of this theory as other reasons for the experimentally observed gradient of proliferation could not be excluded.

## Discussion and Outlook

In this study we have developed a computational model that was able to quantitatively explain spatial-temporal data on regeneration of the liver after partial hepatectomy (PHx) in mouse occurring by growth of the remnant non-dissected lobules until the liver mass prior to dissection had been recovered.

The computational model considered a four-cells-thick tissue slice of an entire lobe at the resolution of an individual hepatocyte. Besides the individual hepatocytes, the model considered the sinusoidal network, central and portal veins. Hepatocytes were mimicked as homogeneous elastic sticky spheres capable of movement, growth and division. The model was built upon an earlier model of regeneration after drug-induced peri-central liver lobule damage [32] but underwent some important modifications. To ensure compatibility between our novel liver lobe model with this earlier liver lobule model, aiming prospectively at a full virtual liver model, most model parameters were kept from that lobule-scale model. The repulsive force between cells, and cells and capsule mimicking the resistance of a cell to compression was modified based on recent findings in multicellular spheroids growing against the mechanical resistance of an elastic alginate capsule, because this case might be considered similar as a growing population of hepatocytes expanding a lobe against the mechanical resistance of the Glisson capsule [51].

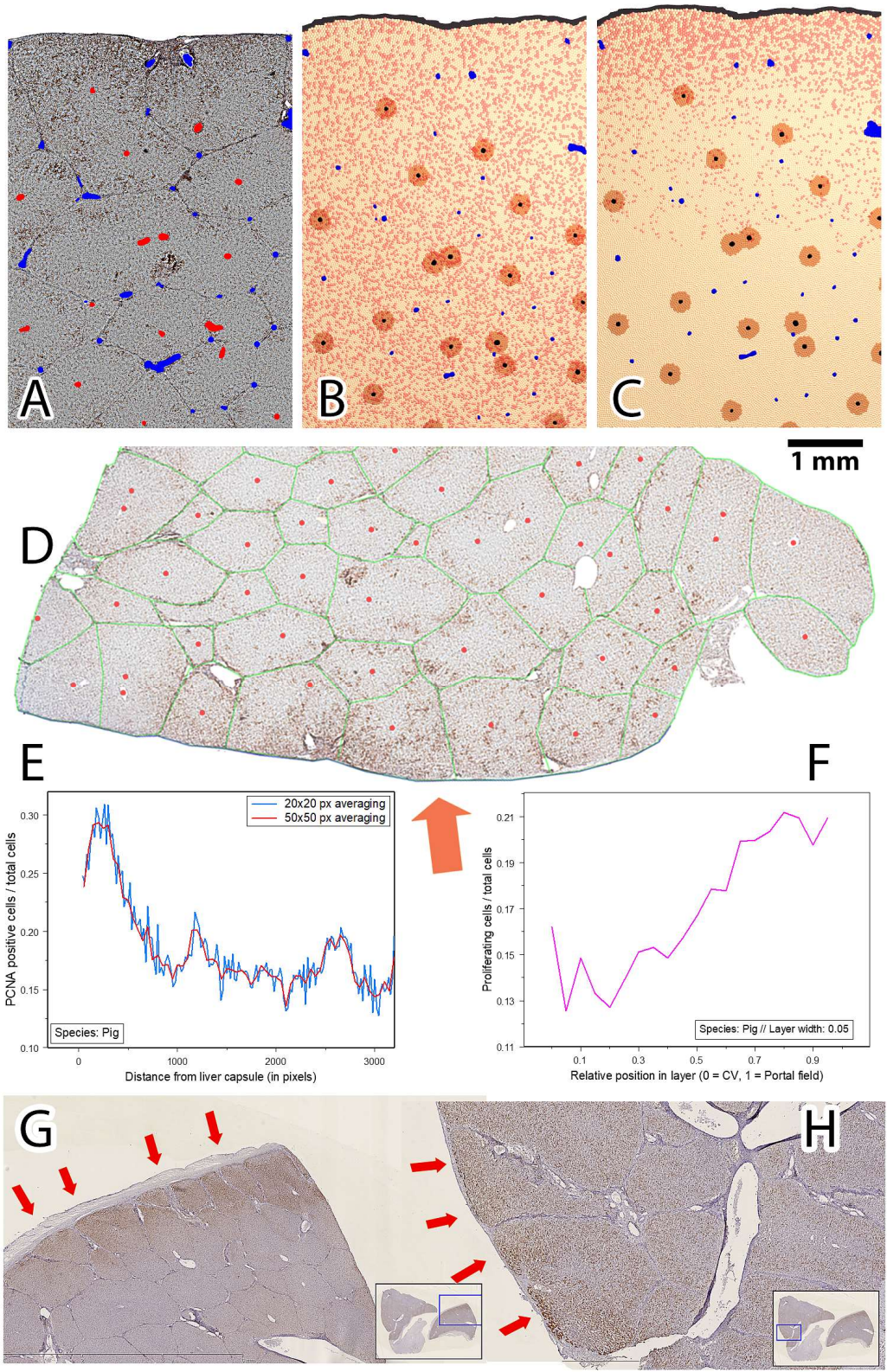

**Figure 7:** (A) Bright field micrograph of pig liver t=2 days after PHx (Red = Central veins detected by image processing and analysis, Blue = portal veins; height of sample: 3.5mm, width: 2.5mm). B) Predicted proliferation scenario in pig during first proliferation wave (t = 1 days) and C) after t = 2 days. (D) PCNA stained micrograph (whole slide scan) of a part of a pig liver 14 days after PHx. Proliferation is mainly localized near the Glisson capsule (orange arrow). (E) (Lower left) Quantification of proliferation within the lobe in relation to the distance to the Glisson capsule. (F) Quantification of proliferation within the lobule shows increased periportal proliferation. Proliferation pattern in further pig livers (2 days after PHx). This experimental data also indicates possible increased proliferation near the Glisson capsule (orange arrows).

In presence of a biomechanical growth control (BGC) that has not been considered in the model of the regenerating liver lobule after drug-induced peri-central lobular damage [32], the model was able to explain the experimentally found doubling of liver lobule size, the proliferation kinetics and the spatial pattern of proliferating cells. BGC assumes cells are only able to progress in the cell cycle if the pressure on them is below a certain threshold. This is consistent with the growth function found in [51]. In absence of BGC the repulsive forces between the cells are insufficient to guarantee a sufficiently fast expansion of the Glisson capsule unless the proliferation is set so high, that the cells reveal unrealistic compressions. Even assuming directed migration of the hepatocytes towards the Glisson capsule turned out for realistic magnitudes of forces to be insufficient to avoid occurring of such unrealistic compressions. In presence of BGC the compressions were significantly reduced. Interestingly, BGC caused a checkerboard-like proliferation pattern indicating that BGC minimized for each proliferating cell the number of its proliferating neighbor cells. This observation suggests that BGC results in a spatial pattern of cell proliferation reminiscent of what is known from activator-inhibitor mechanisms with short range activator and long range inhibitors [39]. This prediction shall be studied in future experiments.

The same model was then tested for pig liver that is much larger than mouse liver. For this purpose, the architectural values of the model have been adapted to values obtained by image analysis in pig. For pig liver the model predicted that cell proliferations after the first wave of proliferation, should be most pronounced close to the Glisson capsule, while in mouse liver, proliferation was homogeneous isotropic over the liver lobe during the entire regeneration period. This finding provided another testable model prediction.

A first pilot experiment indicates that this prediction might be correct even though a careful study with more pigs would be necessary to validate this prediction. The cause of the difference in the proliferation pattern for small and large liver lobes is that interior proliferating cells in large lobes have to shift much more material in order to generate space for division than this is the case in small liver lobes. This leads to the buildup of a pressure gradient with highest pressure values in the center of the lobe and lowest pressure values close to the Glisson capsule. BGC limits the pressure value in the lobe interior but does not avoid the gradient at the border.

The presence of a pressure gradient with lowest pressure at the border is a generic feature unless the Glisson capsule is so stiff, that no expansion of the lobe by cell division would be possible anymore. This can be seen by considering the two hypothetical limits of infinitely stiff versus soft Glisson capsule. In the theoretical limit where the Glisson capsule would be infinitely stiff the pressure would be homogeneous and isotropic with no gradient (as in a pressure cooker). Due to the inextensibility of the capsule in that case either no net change of the cell population size can occur, or the cells would have to shrink, or to occupy space normally taken by sinusoids or ECM. This limit does not apply in the regenerating liver after PHx. The other hypothetical limit is that with infinitely soft (or no) Glisson capsule. In that case the cells close to the Glisson capsule behave almost as cells at an interface to a liquid medium hence a pressure gradient develops with low pressure at the border and increasing pressure towards the interior. This behavior is reminiscent to growing 3D spheroids and monolayers where a proliferating layer forms, which in monolayers is by construction not nutrient-controlled, and in multicellular spheroids is not nutrient controlled if sufficient nutrients are available [20]. The case of the regenerating liver must be closer to the latter case.

We concluded that the Glisson capsule can be expanded by proliferation of the enclosed cell population only if the stiffness of the capsule is sufficiently moderate. In this case a pressure gradient forms. The gradient is the more pronounced, the lower is the stiffness of the Glisson capsule. If the pressure (or compression, as in [51]) remains on average over the entire lobe below the threshold values at which cell cycle progression would be inhibited, cell proliferation is homogeneous and isotropic. This could be the case in mouse. If at some region in the lobe interior the pressure is larger than the threshold value, regions with more (border) and less (interior) cell proliferations form. In this case the growth is slower than in the former case. This could be the case in pig. If, as in mouse, cells proliferate homogeneously and isotropically in the lobe one might expect exponential growth of the cell population enclosed by the Glisson capsule, but this is unlikely to be detectable as (1) only a fraction of cells enter the cell cycle and as (2) the lobe only increases its cell population size by a factor of roughly three, (3) the Glisson capsule still provides a resistance that might depend on the degree of its extension, and (4) the density of cells might slightly increase.

Still it is possible that we might have under-or overestimated the mechanical resistance of the Glisson capsule, or the threshold pressure for proliferation, which are not precisely known. However, in the case of a stiffer capsule the results are unaltered as long as the threshold pressure is larger than the pressure that forms inside the lobe. If the latter is not the case, elevating the pressure threshold would again result in homogeneous, isotropic growth (SI-Fig. 1).

We here did not consider the possibility that the Glisson capsule itself grows. Long-term one would certainly expect that remodeling of the capsule after or during expansion of the lobe would relax the tensile stress in the Glisson capsule, and might even put it to zero i.e., as if there were not Glisson capsule. This however would not be expected to alter the conclusion of this work as even in total absence of a Glisson capsule stress would build up in the interior of the lobule and compress cells un-physiologically in absence of BGC as can be seen by comparison with simulations of growing monolayers or multicellular spheroids, where an enclosing capsule does not exist and where the diameters are of the same order of magnitude as for a mouse lobe.

Besides regeneration of a liver lobe after partial hepatectomy BGC is compatible with regeneration of a peri-central drug-induced damage as we finally demonstrated by simulations of this process in presence of BGC. BGC is fully compatible with active migration towards the central necrosis and hepatocyte-sinusoid-alignment during that regeneration process (SI). Hence the conclusions of that earlier work in [32] remain unaltered.

## Acknowledgement

DD gratefully acknowledges support by the projects EU-PASSPORT, ANR-IFLOW, ANR-iLITE. BMBF-LiSym, and BMBF-Lebersimulator. SH gratefully acknowledges support by BMBF-LiSyM and DFG (HO 4772/1-1).

# Supplemental Information (SI)

## Varying mitotic index and threshold pressure for cell cycle progression.

In this section, we analyze the impact of different model parameters on lobe growth.

Increasing the mitotic index defined as the probability of a cell to start proliferating within 24h (the experimentally found values was 0.5) increased the lobe in the simulation, but also led to small undulations at the Glisson capsule (shown for t = 3 days in SI-Fig.1,A-D, SI-Fig.1I) reminiscent of buckling observed in the basal layer of skin or oral mucosa or in irradiated crypts [23]. Buckling occurs when the stabilizing shear stress (or bending) force is outcompeted by the destabilizing cell proliferation [19]. The more rigid the Glisson capsule was, the more unlikely was buckling to occur at a certain lobe size. This effect might be tested by increasing the cell cycle progression rate. In the corresponding simulations we chose E = 5.5 kPa for the Glisson capsule.

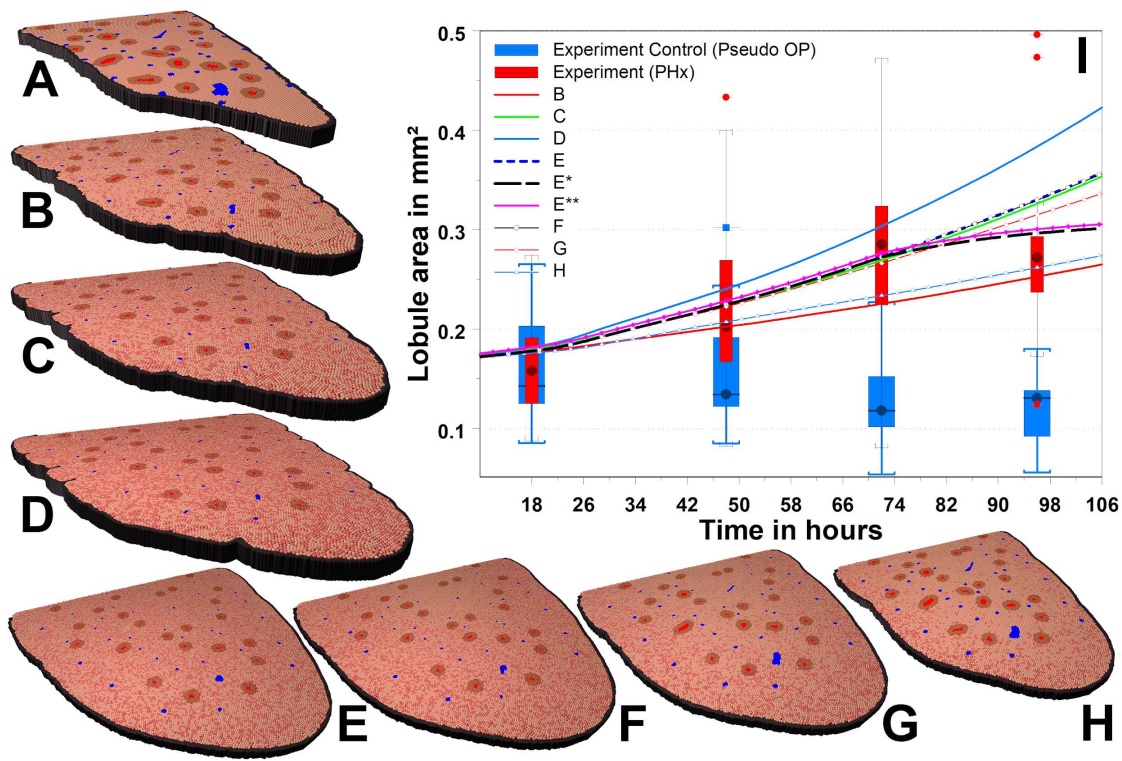

**SI-Figure 1**: A) Model of 4 cell layers height constructed from experimental images (see Fig.1) at t=0 days (initial state). B-D) Prediction of liver lobe pattern for different mitotic indices (achieved by varying pressure threshold) at t=3 days. Mitotic index: B) 0.3, C) 0.5 (as found in the experiments) and D) 0.8. Mitotic index = probability of a cell to start proliferating within 24h. Young modulus of Glisson capsule (grey color): 5500 Pa. (E-H) Model predictions for different proliferation inhibition thresholds w after t = 5 days: E) w=300 Pa (leads to a homogeneous distribution of proliferation as found in the experiments, see Fig.3), F) w =200 Pa, G) w = 150 Pa, H) w = 100 Pa. Furthermore, in these simulations the Young-modulus of the capsule was chosen E = 20 kPa. In comparison to A)-D) this led to a smoother Glisson capsule with no undulations. Only one half of the lobe was shown but all simulations were carried out for whole lobes. I) Model kinetics of B)-H) carried out to study the effect of the mitotic index and pressure-based inhibition of proliferation (E*). In the curves E* (black, dashed line) and E** (magenta, solid line) the simulations were carried out with a global inhibition of proliferation once the original liver cell population size has been restored. Therefore, this simulation saturates after 3-4 days while the other simulations do not. In the curve E** (magenta, solid line), the Young's modulus was increased by the factor 30 for small cell-cell distances to study the effect of various cell compressibility.

In the computer simulations, the fraction of cells entering the cell cycle also changed if the pressure threshold, at which a cell would exit the cell cycle, was modified (SI-Fig.1E-H). A lower pressure threshold results in a slower lobule growth speed (SI-Fig.1I) but only for w<200Pa (reference: 300Pa) the deviation was clearly detectable with a half as big lobule increase at day three. The reason is that below w=200Pa the pressure for interior cells is above the proliferation threshold resulting in only growth close to the Glisson capsule (SI-Fig.1G/H).

## Liver regeneration after overdose of drugs generating pericentral liver lobule damage

In [32] we had studied liver regeneration after intoxication with $CCl_4$. Intoxication by $CCl_4$ leads to a pericentral necrotic lesion, that is then closed within a regeneration process taking about one week to restore the liver mass and 2 weeks to restore liver microarchitecture. The final model obtained (here referred to as (sub-) model 3) out of three alternative models studied in [32] corresponds to the model detailed in "Material and Methods", except that BGC was not considered in any of the three models, and that the repulsive forces at large compression were smaller. In each of the models we had directly used the experimentally determined spatial-temporal proliferation pattern after $CCl_4$ administration as a model input by sampling from the spatial-temporal distribution of BrdU-positive cells to select cells that enter the cell cycle in the computer simulation. The question of this section is to evaluate if including BGC would have modified the results and conclusion of the model of liver regeneration after $CCl_4$ administration, which could be critical as that model formed the basis of the model at the lobe scale in this work.

First, we briefly summarize how we quantitatively compared them to data and the three models 1-3, before we present the simulation simulations with the updated versions of the models that include BGC.

Process parameters: In order to evaluate the agreement of the results of each submodel with the experimental findings, we had considered three "process" parameters (PPs, [22, 32]) that we equally measured in experimental images as in the spatial multicellular tissue configurations, namely, (i) the number of hepatocytes per lobule area (PP1), (ii) the area of the necrotic lesion (PP2) and (iii) the hepatocyte-sinusoid interface area fraction (PP3) as a measure for the regeneration of liver lobule microarchitecture. Within a simulated sensitivity analysis, we had varied each model parameter within its physiological range to identify the best possible match between that model and the data. The physiological range could well be identified as the model was parameterized by measurable, meaningful biophysical and bio-kinetic parameters [22].

Models: In "(sub-)model 1" we had assumed that micro-motility and cell division are both isotropic. This model was not able to close the necrotic peri-central lesion within the experimentally observed regeneration time (i.e. the process parameter "necrotic area", PP2 did not drop to zero in time) but it was able to regenerate the number of cells per lobule (process parameter PP1) by generating a population of strongly compressed hepatocytes at the lobule periphery. In an improved "(sub-)model 2", we had considered directed cell migration modeled as biased micro-motility in the direction of the necrotic lesion. Even though this model was able to explain the restoration of the liver cell number (PP1) and the closure of the necrotic lesion (PP2) within the experimentally measured time, it could not explain the regeneration of liver architecture (PP3). Only a further improved "(sub-)model 3" in which a novel mechanism, HSA, the alignment of hepatocytes along local sinusoidal vessels after cell division had been introduced (Fig.2C), was able to explain the experimentally observed regeneration

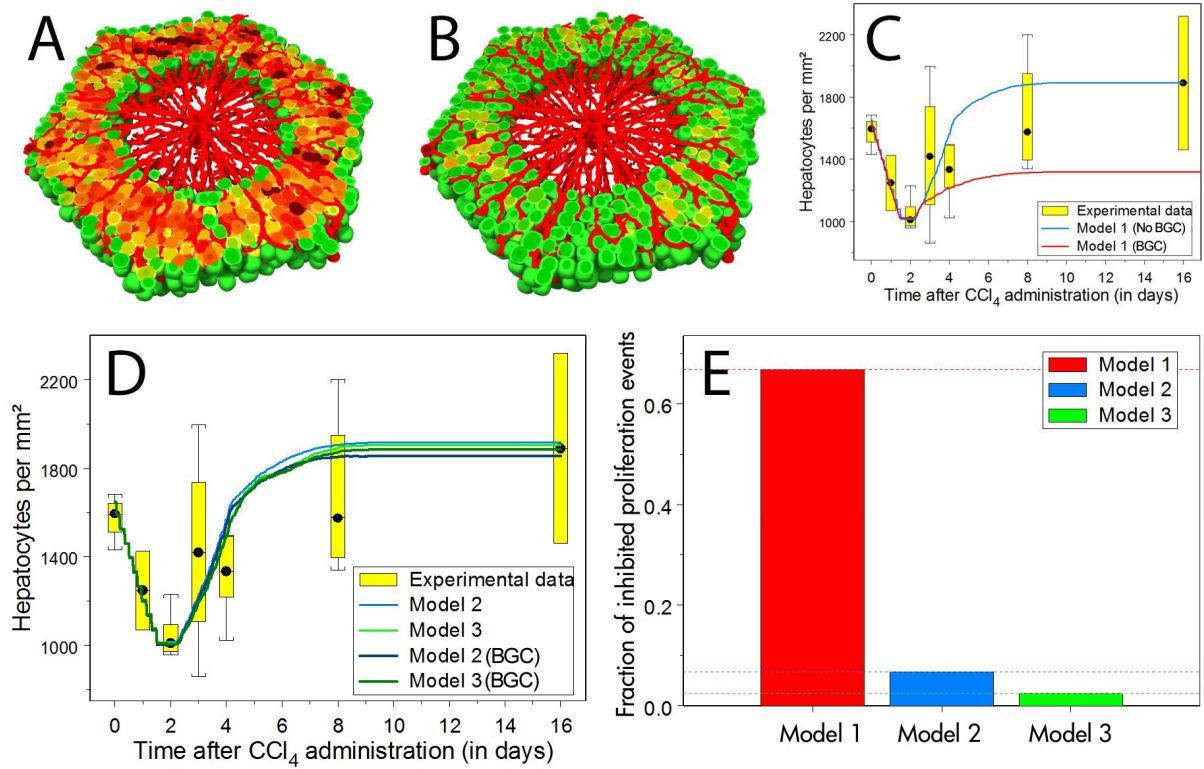

**SI-Figure 2:** (A) Cell volume during regeneration after $CCl_4$ intoxication if unrealistic compression was not inhibited. (Green = the volume of an isolated sphere, Red = 0.3 times that volume) (B) The same simulation only with the assumption that entrance into the cell cycle is possible only, if the local pressure does not overcome a critical threshold. A Voronoi-based space partitioning analysis has been used to approximate the volume of the hepatocytes in the model. (C) Time course of cell population size per liver lobule in cases (A) and (B). (D): Time course of cell population size per liver lobule in models 2 and 3 with and without pressure inhibition by proliferation (BGC). (E): Fraction of cases in which proliferation impulses do not lead to proliferation due to pressure-controlled growth inhibition for models 1, 2 and 3 showing that the presence of a mechanism that inhibits proliferation by pressure (BGC) does not modify the regeneration dynamics of models 2 and 3 in [32].

process i.e., all parameters PP1-PP3. In order to ensure that all physiological parameters would be captured, a simulated sensitivity analysis has been pursued, varying each model parameter within its physiological range and quantifying the deviation of each process parameter PP1-PP3 between data and simulation.

To compute the potential impact of BGC-based cell cycle entrance on those results, we re-simulated all (sub-)models 1-3 in [32] now including the pressure-based cell cycle progression mechanism BGC as explained in (A5(B)) i.e., we first use the experimental proliferation pattern to pick cells as candidates for cell cycle progression (A5(A)) , and then choose those cells to enter the cell cycle for which the pressure is below a threshold value (A5(B)). Vice-versa, cell cycle entrance was inhibited if the local pressure overcame a critical threshold. Consequently, re-running the simulations with model 1 extended by a pressure-controlled growth inhibition removed unphysiological compression of hepatocytes (SI-Fig.2B). However, as a consequence model 1 was then not be able anymore to retain the experimentally observed time development of the hepatocyte population within each lobule (PP1, SI-Fig.2C, red line) as approximately two thirds of all proliferation events were suppressed (SI-Fig.2E). Increasing the number of cells chosen under step (A5(A)) as candidates in model 1 would not have much changed the fraction of cells entering the cell cycle as those would be rejected due to the too high pressure in the 2nd step (A5(B)) so the too small number of cell cycle progression events for (sub-)model 1 cannot be balanced by an increase of candidates under step (A5(A)).

Adding the pressure-controlled growth inhibition to (sub-)models 2 and 3 had almost no effect on the regeneration kinetics since a non-physiological compression was already avoided due to active cell migration towards the necrotic lesion (SI-Fig.2C/D). In both model variants, the introduction of a pressure-controlled growth inhibition suppressed only a minor fraction of proliferation events (SI-Fig.2E, model 2: ~6%, model 3: ~3%). We focused only on PP1 as in these parameters the differences showed up.

## Regeneration simulations varying lobule thickness and other parameters

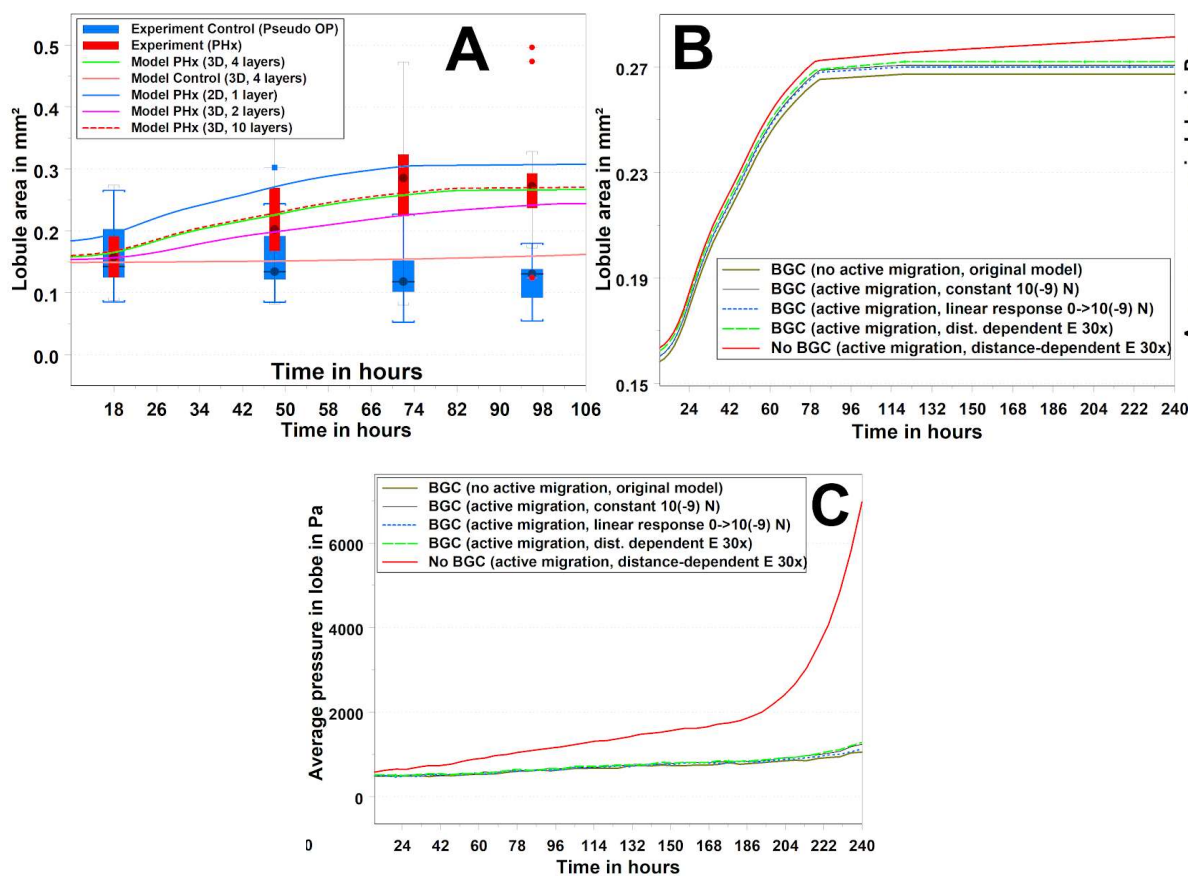

**SI Figure 3**: (A) Regeneration vs thickness (z-height) of the simulated lobes (1, 2, 4, 10 cells). (B) Comparison of area increase for different regeneration mechanisms (brown: original model, no active migration towards the Glisson capsule; black, blue, green: active migration towards the Glisson capsule, with constant micromotility force of $10^{-9}$N (black), varying migration force (blue), distant-dependent cell Young modulus to mimic the strong repulsive force upon large cell compression (green), absence of BGC but with strong repulsive force upon large cell compression (red). (C) Comparison of average pressure in lobe for the models shown in (B).

## Simulated BrdU proliferating pattern

We here for complement present the simulated BrdU pattern for regeneration after PHx with BGC. BrdU only stains cells in the S-phase, which is about 8h long. In contrast, Fig. 5A/B show all cells in the cell cycle corresponding to a simulated KI-67 proliferation pattern. In that case the BGC generated specific "checkerboard" proliferation pattern (Fig.5A/B) could be identified as depicted in the histogram (Fig.5C). Now (in SI-Fig.4) we simulated an experimental standard protocol where mice were sacrificed 2 hours after injection of BrdU. In that case, those cells that were in the S-phase during the injection and those cells that entered the S-phase within the two hours until the mice were sacrificed were BrdU-stained. The simulation results of the histogram equivalent to that in Fig.5C/D for KI-67 staining now shows only minor differences demonstrating that labeling S-phase only is insufficient to detect the BGC-generated specific growth pattern.

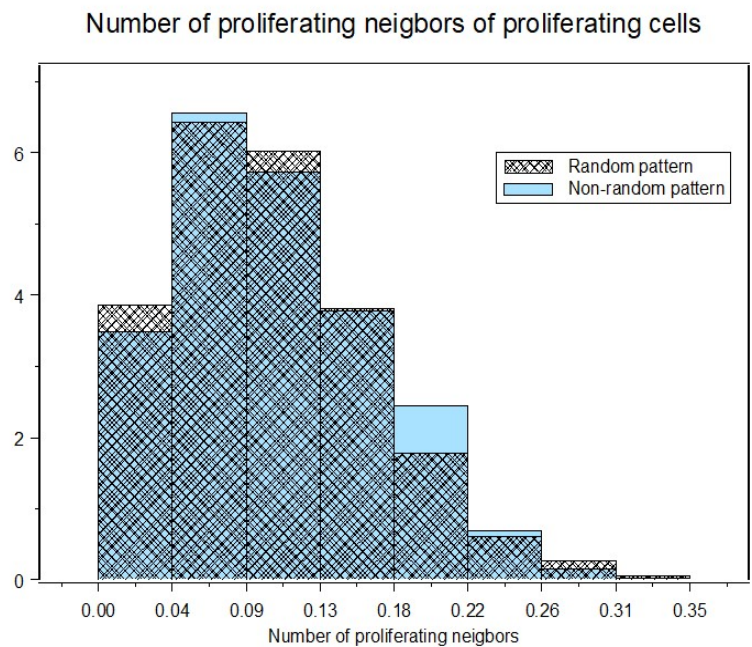

**SI-Figure 4**: Frequency histograms for the number of proliferating cells in the vicinity of a proliferating cell predicted by the model for BrdU staining (non-random pattern demarcates the BGC mechanism). Compare to Fig.5C for the model prediction in case of an idealized KI-staining.